\documentclass[journal]{IEEEtran}
\usepackage[english]{babel} 
\usepackage{amsmath,amssymb,amsfonts}
\usepackage{graphicx}
\usepackage{caption}
\usepackage{booktabs}
\usepackage{xcolor}
\usepackage{physics}
\usepackage{tikz}
\usepackage{pgfplots}
\pgfplotsset{compat=1.18}
\usepackage{comment}
\usepackage{blindtext}
\usepackage[ruled,vlined]{algorithm2e}
\usepackage{xcolor}
\usepackage{soul}
\SetKw{KwBy}{by}
\SetKw{KwTo}{to}
\SetKwComment{Comment}{/* }{ */}
\usepackage{float}
\usepackage{balance}
\usepackage{ftnxtra}
\usepackage{fnpos}
\usepackage{subcaption}
\usepackage{hyperref}
\usetikzlibrary{shapes.geometric, arrows}

\tikzstyle{startstop} = [rectangle, rounded corners, minimum width=0.3\columnwidth, minimum height=1cm,text centered, draw=white, fill=white]

\tikzstyle{process} = [rectangle, minimum width=0.4\columnwidth, text width = 0.4\columnwidth, minimum height=1cm, text centered, draw=black, fill=orange!60]

\tikzstyle{process2} = [rectangle, minimum width=0.2\columnwidth, text width = 0.2\columnwidth, minimum height=1cm, text centered, draw=black, fill=orange!30]

\tikzstyle{arrow} = [thick,->,>=stealth]

\tikzstyle{textbox} = [rectangle, rounded corners, minimum width=0.1\columnwidth, minimum height=1cm,text centered, draw=white, fill=white]

\tikzstyle{inout} = [rectangle, rounded corners, minimum width=0.1\columnwidth, text width = 0.05\columnwidth, minimum height=1cm,text centered, draw=black, fill=orange!60]

\tikzstyle{gate} = [rectangle, rounded corners, minimum width=0.05\columnwidth, text width = 0.1\columnwidth, minimum height=1cm,text centered, draw=black, fill=white]

\tikzstyle{hidden} = [rectangle, rounded corners, minimum width=0.1\columnwidth, text width = 0.05\columnwidth, minimum height=1cm,text centered, draw=black, fill=red!60]

\tikzstyle{met} = [circle, minimum width=0.05\columnwidth, text width = 0.05\columnwidth, minimum height=1cm,text centered, draw=black, fill=white]

\begin{document}
\bstctlcite{IEEEexample:BSTcontrol}

\title{CryoNet: A Deep Learning Framework for Multi-Modal Debris-Covered Glacier Mapping. \\A Case Study of the Poiqu Basin, Central Himalaya}
\author{Farzaneh Barzegar,~\IEEEmembership{Student Member,~IEEE,}~{Tobias Bolch},~{Norbert Kuehtreiber}~and~Silvia~L.~Ullo,~\IEEEmembership{Senior~Member,~IEEE}

\thanks{S. L. Ullo is with the Engineering Department, University of Sannio, Benevento, Italy, email: ullo@unisannio.it; 
F. Barzegar, N. Kuehtreiber, and Tobias Bolch are with the Institute of Geodesy, Graz University of Technology, Graz, Austria, email: farzaneh.barzegar@tugraz.at, norbert.kuehtreiber@tugraz.at, tobias.bolch@tugraz.at}}
%
\maketitle
\begin{abstract}

Glaciers play a critical role as freshwater reserves and indicators of climate change, yet their automatic delineation, especially for debris-covered glaciers, remains challenging due to spectral similarity with surrounding terrain. This study introduces CryoNet, a deep learning framework that leverages a rich multi-modal dataset combining Sentinel-2 optical imagery, DEM-derived topographic variables, spectral indices, Principal Component Analysis (PCA), InSAR coherence and phase, tasseled-cap features, and GLCM texture to discriminate clean-ice glaciers, debris-covered glaciers, and glacial lakes. CryoNet is an encoder–decoder CNN with nested skip connections and spatial–channel Squeeze-and-Excitation (scSE) attention, built upon a ResNet101 encoder to capture hierarchical contextual and spatial features. The study is conducted in the Poiqu Basin in the central Himalaya, and transferability is evaluated by applying the trained model to the Mont Blanc Massif in the Alps. We additionally analyse the importance of each data layer in improving glacier mapping performance. The proposed model achieves an overall IoU of 90.52\%, mean Recall of 98.08\%, and mean Precision of 92.26\%. For debris-covered glaciers specifically, CryoNet obtains an IoU of 90.46\%, a recall of 95.79\%, and a precision of 94.21\%. Across both per-class and overall metrics, CryoNet surpasses DeepLabV3+, SegFormer, and U-Net, taken as state-of-the-art (SOTA) references, demonstrating its effectiveness for robust glacier mapping in complex high-mountain environments.

\end{abstract}

\section{Introduction}

\IEEEPARstart{G}{laciers} play a vital role in the Earth system dynamics as crucial sources of freshwater~\cite{immerzeel2020importance} and as sensitive indicators of climate change, making them reliable proxies for global warming~\cite{bojinski2014concept}. Their ongoing shrinkage and development represent some of the fastest ecosystem shifts worldwide ~\cite{Bosson2023}, with far-reaching consequences such as contributing to global sea-level rise~\cite{portner2019ocean}. Consequently, continuous monitoring of glaciers is essential for understanding changes in glacier mass and runoff, all of which require accurate and up-to-date glacier inventories as a fundamental prerequisite~\cite{Pfeffer2014}. 

Debris-covered glaciers are defined by (Kirkbride, M. P. et al. 2011)~\cite{kirkbride2011debris} as ‘a glacier where part of the ablation zone has a continuous cover of supraglacial debris across its full width’. They are common in several mountain ranges and dominant in the Himalayas~\cite{Racoviteanu2012}, ~\cite{Herreid2020}, and other regions as well. Studying debris-covered glaciers is important as they are prone to the formation of hazardous glacial lakes, and their response to climate change is more complex than that of clean-ice glaciers ~\cite{Benn2012}. Therefore, their continuous monitoring is important.

Accurate delineation of glaciers (debris-covered and clean-ice) is the first and vital step of any task in glacier monitoring. There are available glacier inventory repositories, namely the GLIMS  (Global Land Ice Measurements from Space) (www.glims.org) and RGI (Randolph Glacier Inventory) ~\cite{RGI7.0_2023}. GLIMS offers extensive data over time compiled by different methods, data, and experts, and RGI is a time-based subset of GLIMS. Despite their importance, due to inconsistent production, existing glacier inventories often suffer from shortcomings, including outdated coverage and data quality ~\cite{Maslov2025}. 
In particular, debris-covered glaciers are just partially categorized separately, and their mapped outlines show even greater inconsistency~\cite{Lippl2018}, ~\cite{Paul2013}. 

Remote sensing (RS) has been a powerful tool in glacier-related studies, as glaciers are located on high mountains, which are not easily accessible regularly, considering the fact that severe weather is also probable. Traditional mapping methods using RS data include manual digitization and spectral band ratio techniques, which present significant limitations. Manual delineation is a tedious task, especially when the area is large. Moreover,  the accuracy depends on the experience of the analyst and the available satellite data. The band ratio method fails in distinguishing debris-covered glaciers from surrounding terrain due to their spectral similarity. While several studies showed that debris-cover can be identified by including topographic parameters, thermal information, and indications for glacier flow (e.g.~\cite{Smith2015}, ~\cite{Mlg2018} and ~\cite{Bolch2007}), the achieved accuracy remained low or needed significant manual improvement. As a result, automatic mapping of glaciers, particularly debris-covered glaciers, remains a considerable challenge. In addition, classical machine learning (ML) techniques such as random forest (RF) have also been widely used for debris-covered glacier mapping  \cite{norova2026systematic}. For example, \cite{ahmad2026multi} applied a Random Forest classifier on Landsat time-series data to analyze multi-decadal changes in debris-covered glaciers in the Zanskar Himalaya. Although ML approaches achieve reasonable classification accuracy, they rely on manually engineered features and lack the ability to automatically learn high-level abstract representations.

Given the limitations of traditional techniques, deep learning (DL) has emerged as a powerful approach for automated glacier delineation. Over the past decade, DL has achieved remarkable success in the RS field~\cite{Ma2019}. DL models can process and integrate heterogeneous data sources, such as optical, thermal, topographic, and Synthetic Aperture Radar (SAR), to exploit their complementary strengths. Above all, in contrast with traditional image segmentation methods, which rely on shallow semantic information~\cite{Li2024}, DL learns hierarchical representations that capture meaningful, complex spectral, spatial, and topographic patterns~\cite{Kazanskiy2025} through automatic extraction of high-level abstract features. In addition, as a powerful tool for handling large datasets, DL enables processing across vast regions, overcoming the limitations of some classical methods with large areas. These capabilities ensure robust discrimination of complex land cover types, such as debris-covered glaciers, and provide a transferable framework for continuous and large-scale glacier monitoring.

The ability of DL techniques to address complex glacier mapping challenges has attracted growing interest. For instance, in~\cite{Maslov2025} the authors introduced the GlaViTU model, a Vision Transformer–U-Net framework that combines optical, InSAR, and thermal data, using GLIMS inventories as ground truth for global glacier mapping. Although this study shows promising results in clean-ice glaciers, its performance drops in detecting debris-covered glaciers, particularly in the High-Mountain Asia (HMA), as it is rich in debris-covered glaciers. Xiang et al. ~\cite{Xiang2025} proposed SAU-Net, a U-Net network enhanced with a simple attention module, for binary glacier–background classification based on Landsat 8 thermal, Sentinel-2, and ALOS DEM data. Further experiments include combining U-Net, FCNN, and DeepLabv3+ with attention mechanisms using high-resolution optical and DEM data~\cite{Yang2024_debris_dl}. ~\cite{diaconu2025dl4gam} introduced DL4GAM, a multi-modal deep learning framework based on U-Net for glacier change monitoring using Sentinel -2 imagery and elevation map. Yet, the focus of none of these papers is on separating debris-covered glaciers from clean-ice.

There have been several studies focusing on mapping debris-covered glaciers.
~\cite{zhang2026} proposed a DL-based framework that integrates a channel attention mechanism and a class-weighted Complex Wavelet Mutual Information (CWMI) loss function. They used optical data from Landsat and Gaofen-6.~\cite{KAUSHIK2025100319} assessed the performance of pretrained geo-foundation models for debris segmentation. In~\cite{Xie2022}  GlacierNet2 is introduced, by following an earlier work, GlacierNet~\cite{Xie2020} by the same authors. A CNN-based framework is developed for mapping debris-covered glaciers by delineating the accumulation zone and the debris-covered ablation zone separately and subsequently combining them. Their approach integrates DeepLabV3+ and GlacierNet and generates glacier outlines. However, the final glacier boundaries were obtained only after a multi-stage post-processing pipeline, including DEM-driven terminus correction and hydrological flow analysis to extend mapping to the accumulation zone. Although effective, this hybrid structure is not end-to-end deep learning and depends heavily on algorithmic post-processing and expert-defined parameters such as threshold selection, by limiting its transferability and scalability to other regions or datasets. Moreover, both networks are computationally demanding and require extensive training data. In~\cite{Yang2024_debrisflow_tibet} the glacier debris flow from 2000 to 2016 in Southeast Tibet, China, is studied. In order to map debris-covered glacier inventories, they used DeepLabv3+ trained on Landsat images. Similarly, in~\cite{Lin2023} Deeplabv3+ for debris-cover mapping is used based on optical data from Landsat 8 and Sentinel 2 in HMA. In~\cite{Khan2023} shallow Convolutional Neural Networks (CNNs) are employed to extract spatial and spectral features separately from Sentinel 2 images, NDVI, and NDWI, and feed them to a Fully Convolutional Network (FCN) for the classification of debris-covered glaciers. To overcome limitations due to data availability, they used a generative adversarial network (GAN) for data augmentation.

Although DL approaches have considerably advanced glacier monitoring and mapping, several challenges remain unaddressed. Existing studies often deploy large and complex models that require extensive training data, leading to poor performance when data are limited. Many studies rely only on restricted data sources, such as optical and DEM, ignoring the potential of diverse data modalities, including SAR and thermal data. These limitations also apply to the use of derived features such as texture, geomorphometric parameters, and spectral indices, which provide valuable cues for discriminat\nobreak ing complex land covers. In addition, there is currently no clear study on which data modalities and derived features contribute most effectively to the delineation of debris-covered glaciers using deep learning methods. Furthermore, existing literature mostly focuses on binary classification, such as either discrimination of glaciers from the background, or classification of glacial lakes, or mapping debris-covered glaciers. Several frameworks rely on post-processing or pre-processing steps that reduce automation and transferability of the trained model. 


Motivated by these limitations, this study proposes Cry\nobreak oNet, a novel DL framework for multi-source glacier mapping, enabling the automatic delineation of clean-ice and debris-covered glaciers using multi-modal open-access data. To this end, the potential of data-driven features such as spectral indices, texture, and geomorphometric variables is also exploited, which have been only partially utilized in previous studies.

The main contributions of the proposed work are as follows:

\begin{itemize}
    \item An encoder-decoder Convolutional Neural Net-
work (CNN) architecture featuring nested skip connections,
enhanced with spatial and channel Squeeze and Excitation
(scSE) attention blocks for semantic segmentation with the en-
coder based on ResNet101, to effectively extract hierarchical contextual features;
\end{itemize}

\begin{itemize}
    \item A multimodal open-source dataset and the corresponding derived features as listed in detail ahead in the paper; 
\end{itemize}

\begin{itemize}
    \item Permutation-based channel importance analysis to understand how much each input band contributed to the performance of the segmentation model, and to evaluate the model sensitivity to the different bands; 
\end{itemize}

 \begin{itemize}
     \item The model’s transferability assessment\\
 \end{itemize}

The remainder of this manuscript is organized as follows: Section~\ref{sec:methodology} describes the methodology, Section~\ref{sec:results} presents the results, Section~\ref{sec:disscussion} discusses the findings and section~\ref{sec:conclusion} concludes the findings and main points.

\section{Methodology}\label{sec:methodology}
The overall workflow includes the selection of the study area, compilation and preprocessing of optical, topographic, and radar-based datasets, and the extraction of derived features such as spectral indices, geomorphometric parameters, and texture measures. 
Subsequently, the model was trained and validated to classify clean-ice glaciers, debris-covered glaciers, water, vegetation, and background classes. 
The methodology section is structured in different paragraphs to easily present the work: Section~\ref{sec:region} describes the study region; Section~\ref{sec:data} introduces the datasets and derived features; Section~\ref{sec:model} details the model architecture, Section~\ref{sec:training} the training strategy, and Section~\ref{sec:evaluate} the evaluation metrics.

\subsection{Study Region}
\label{sec:region}

\begin{figure}[t]
\centering
\begin{tikzpicture}
    \node[anchor=south west, inner sep=0] (main) at (0,0)
        {\includegraphics[width=1\columnwidth]{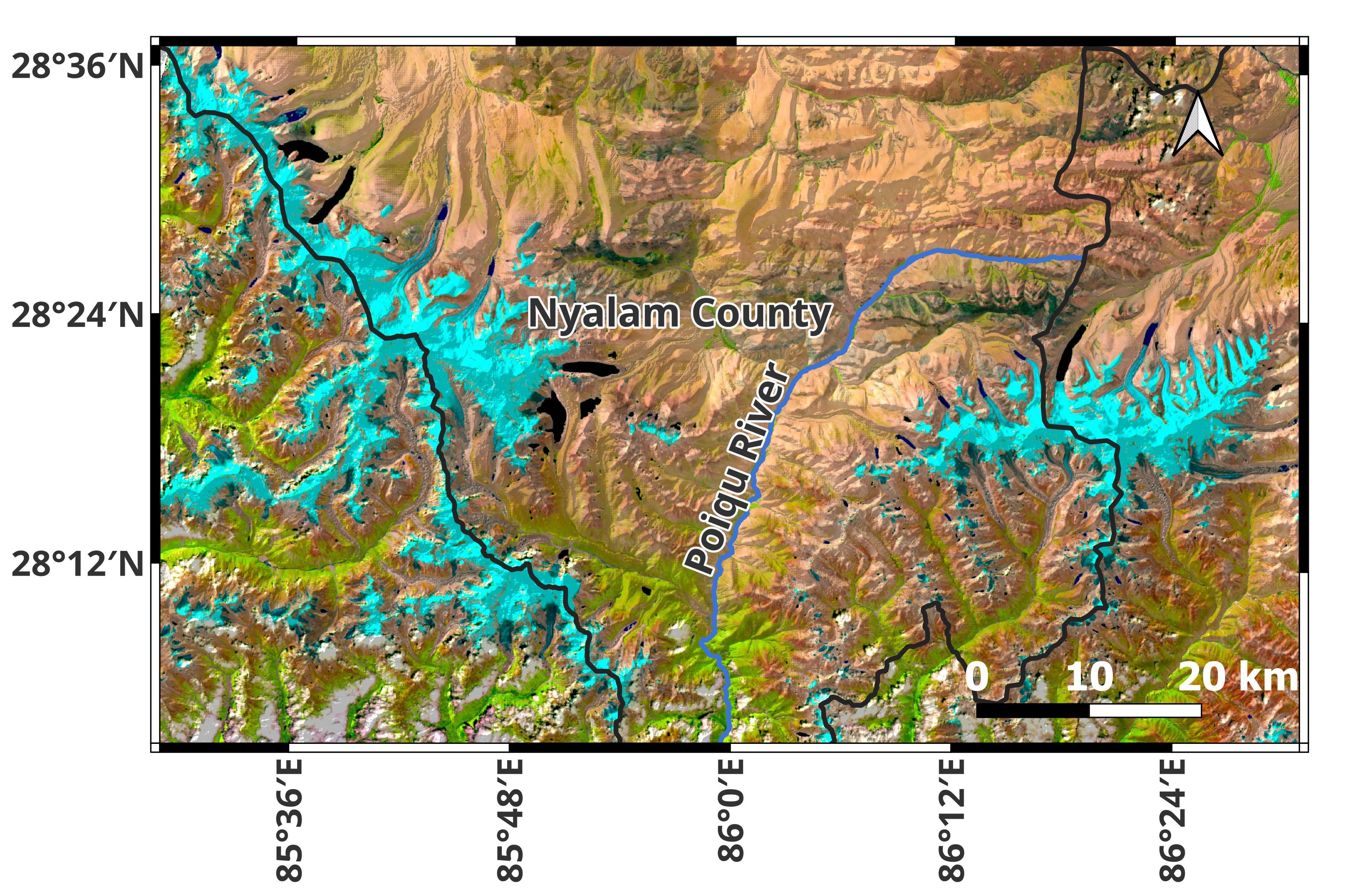}};
    \begin{scope}[x={(main.south east)}, y={(main.north west)}]
        \node[anchor=south west, inner sep=0] at (0.30, 0.83)
            {\includegraphics[width=0.5\columnwidth]
            {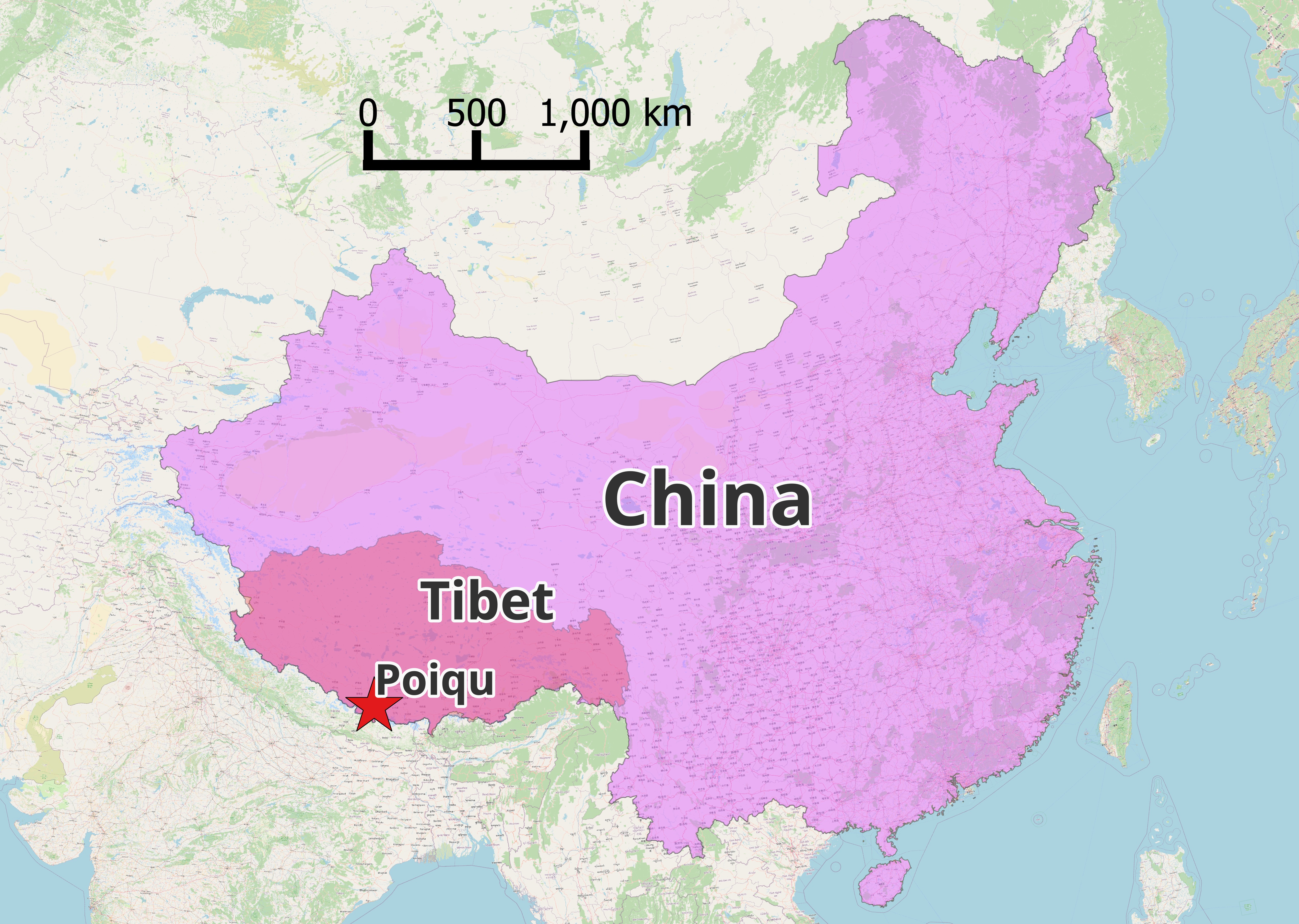}};
    \end{scope}
\end{tikzpicture}
\caption{Poiqu basin in WGS84 geographic coordinate system. The background image: Sentinel 2 acquired on 15 October 2018}
\label{fig:poiqu_basin}
\end{figure}

The main study area is the Poiqu River Basin, located in the central Himalaya, draining southward from the Tibetan Plateau with a relatively high mean elevation to Nepal (Fig.~\ref{fig:poiqu_basin}). Covering an area of approximately 2,000 km², the basin is rich in glaciers and glacial lakes~\cite{Xiang2014}. Recent studies have illustrated that deglaciation has accelerated over the past decades in the Himalaya and that debris-covered glaciers experience comparable or even greater ice loss than clean-ice glaciers~\cite{Maurer2019}, ~\cite{Bolch2022}, ~\cite{bhattacharya2021high}. The presence of abundant glaciers, glacial lakes, and their rapid changes makes the region prone to glacier lake outburst floods~\cite{Wang2024}, ~\cite{zhang2019glacial}, ~\cite{allen2019potentially}, and a region of interest for studying the impacts of climate change.

\subsection{Datasets and derived features}
\label{sec:data}
We utilized multimodal open-source datasets and the corresponding derived features. All data layers were resampled to a spatial resolution of 10~m for consistency. In the remainder of this section, the datasets are described as follows: 

\textbf{Optical imagery:} Sentinel-2 L2A data were used as the primary optical source. The imagery provides 12 spectral bands with spatial resolutions ranging from 10 m to 60 m, all of which were used as input features. In addition, several spectral indices were derived from this dataset. We used data acquired on 15 October 2018. This acquisition date was selected due to its minimal cloud coverage and seasonal snow cover.

\textbf{Digital Elevation Model (DEM): }The Copernicus DEM GLO-30 (30 m resolution) was employed as a source of topographic information and to derive geomorphometric parameters, including slope and aspect. The dataset is provided by the European Space Agency under the Copernicus programme and is available at https://doi.org/10.5270/ESA-c5d3d65.

\textbf{Spectral indices: }

\begin{itemize}
    \item \textit{Normalized Difference Vegetation Index (NDVI)} helps discriminate debris-covered glaciers, as they have no or very sparse vegetation ~\cite{Taschner2002}. It can be calculated using equation~\ref{equ:ndvi}.
\begin{equation}
\centering
\text{NDVI} = \frac{NIR - Red}{NIR + Red}
\label{equ:ndvi}
\end{equation}
    \item \textit{Normalized Difference Snow Index (NDSI)} is a stable band ratio index that effectively detects snow and clean ice while minimizing illumination effects~\cite{racoviteanu2008decadal}. It discriminates debris from snow and ice and is commonly used in the generation of glacier inventories. NDSI can be calculated using equation~\ref{equ:ndsi}.    
\begin{equation}
\centering
\text{NDSI} = \frac{Green - SWIR}{Green + SWIR}
\label{equ:ndsi}
\end{equation}
    \item \textit{Normalized Difference Water Index (NDWI)}helps water distinguishment ~\cite{huggel2002remote}. In addition, since NDSI may misclassify water as snow or ice, NDWI helps mitigate this limitation. It can be calculated using equation~\ref{equ:ndwi}.
\begin{equation}
\centering
\text{NDWI} = \frac{Green - NIR}{Green + NIR}
\label{equ:ndwi}
\end{equation}
    \item \textit{Normalized Difference Glacier Index (NDGI)} introduced by ~\cite{Keshri2009}  can be used for the discrimination of snow and ice versus ice-mixed-debris. Although it is not widely used, it can be particularly useful when applied within predefined glacier outlines as a mask. It can be calculated using equation~\ref{equ:ndgi}.

\begin{equation}
\centering
\text{NDGI} = \frac{Green - Red}{Green + Red}
\label{equ:ndgi}
\end{equation}
 \end{itemize}
 
\begin{figure}[!t]
\centering
    \begin{subfigure}{0.49\columnwidth}
        \includegraphics[width=\linewidth]{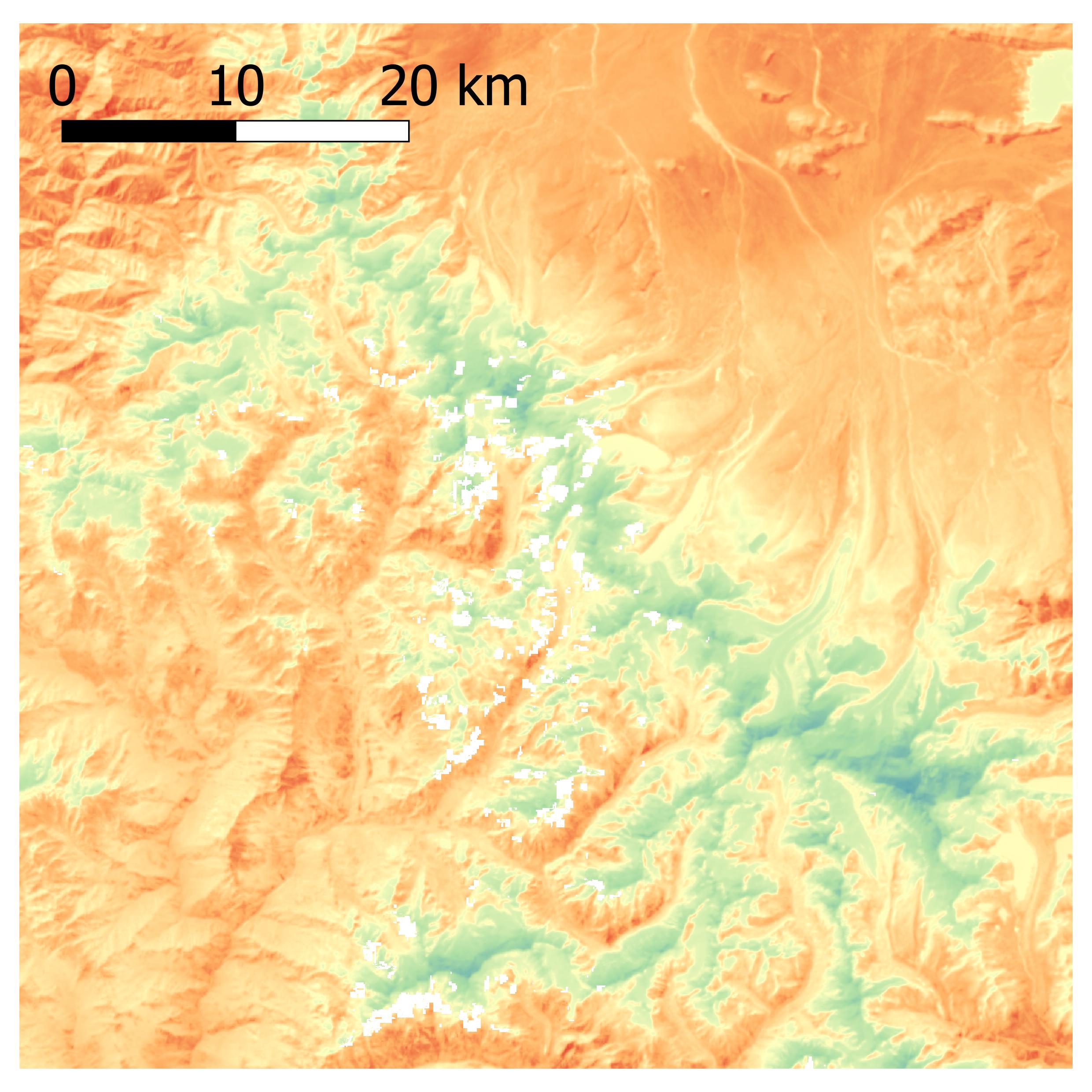}
        \caption{LST}
        \label{fig:LST_GL_bound}
    \end{subfigure}
    \hfill
    \begin{subfigure}{0.49\columnwidth}
        \includegraphics[width=\linewidth]{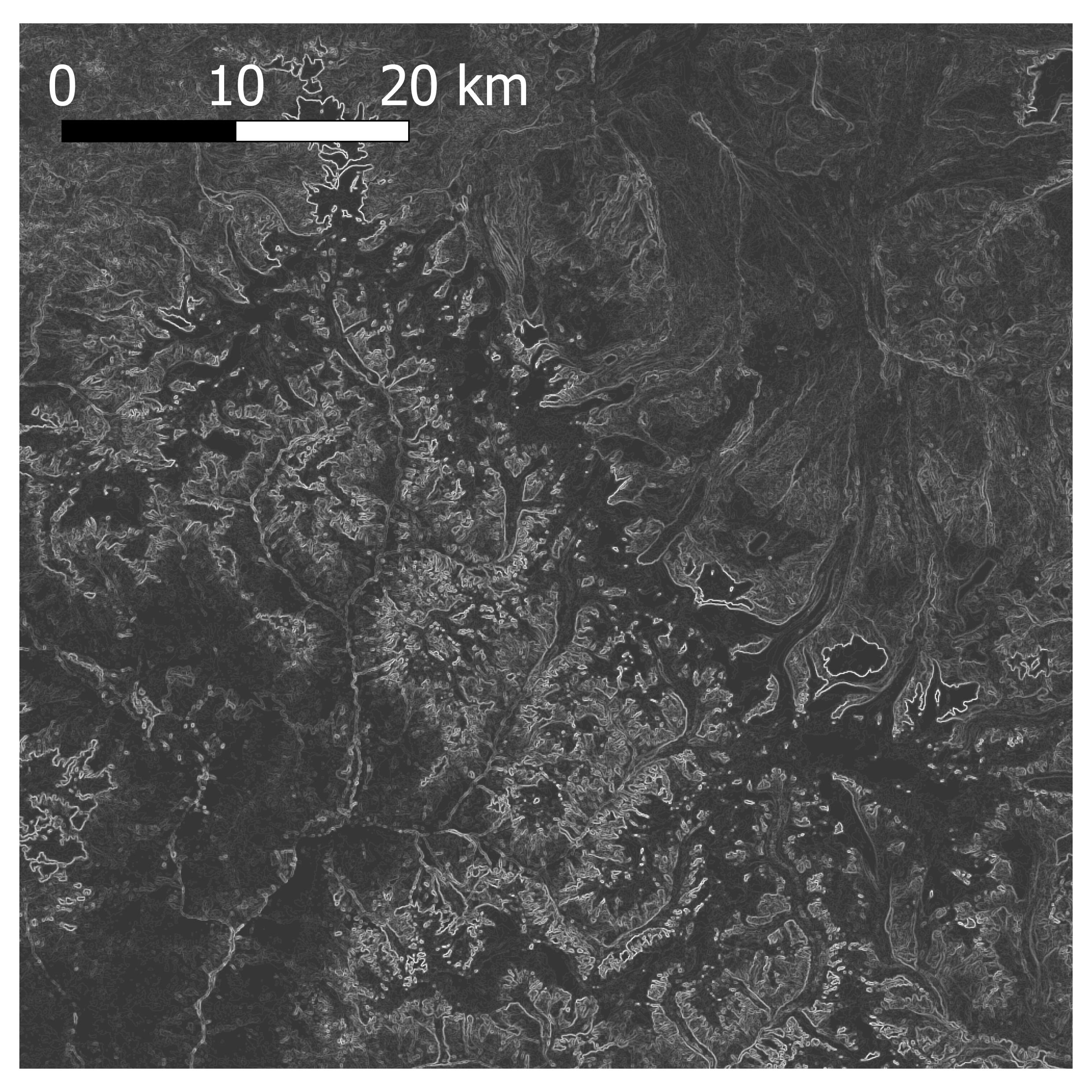}
        \caption{GLCM}
        \label{fig2:glcm}
    \end{subfigure}

    \vspace{2mm} 

    \begin{subfigure}{0.49\columnwidth}
        \includegraphics[width=\linewidth]{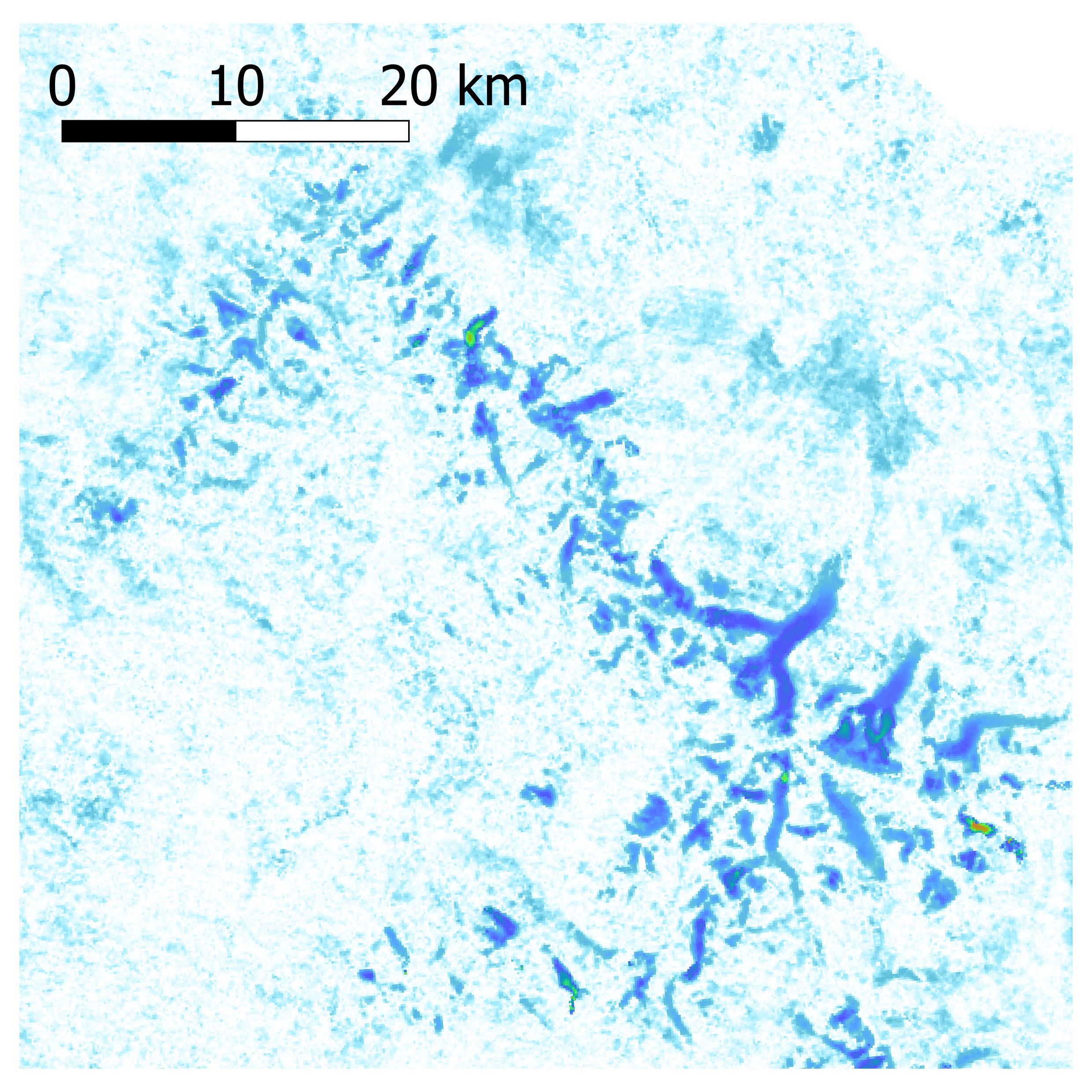}
        \caption{ITS\_live}
        \label{fig:ITS}
    \end{subfigure}
    \hfill
    \begin{subfigure}{0.49\columnwidth}
        \includegraphics[width=\linewidth]{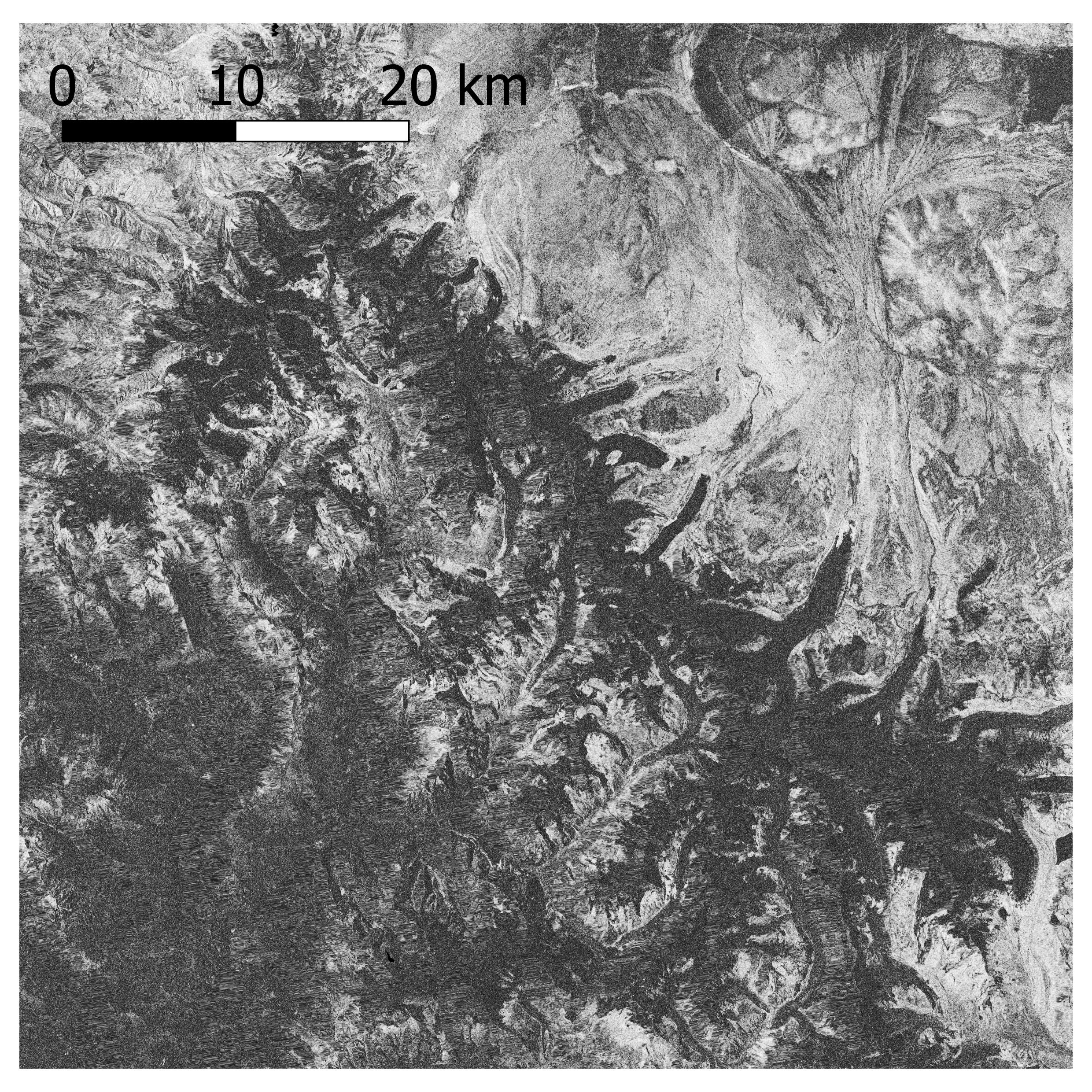}
        \caption{InSAR coherence}
        \label{fig:insar_co}
    \end{subfigure}

    \vspace{2mm} 
    \begin{subfigure}{0.49\columnwidth}
        \includegraphics[width=\linewidth]{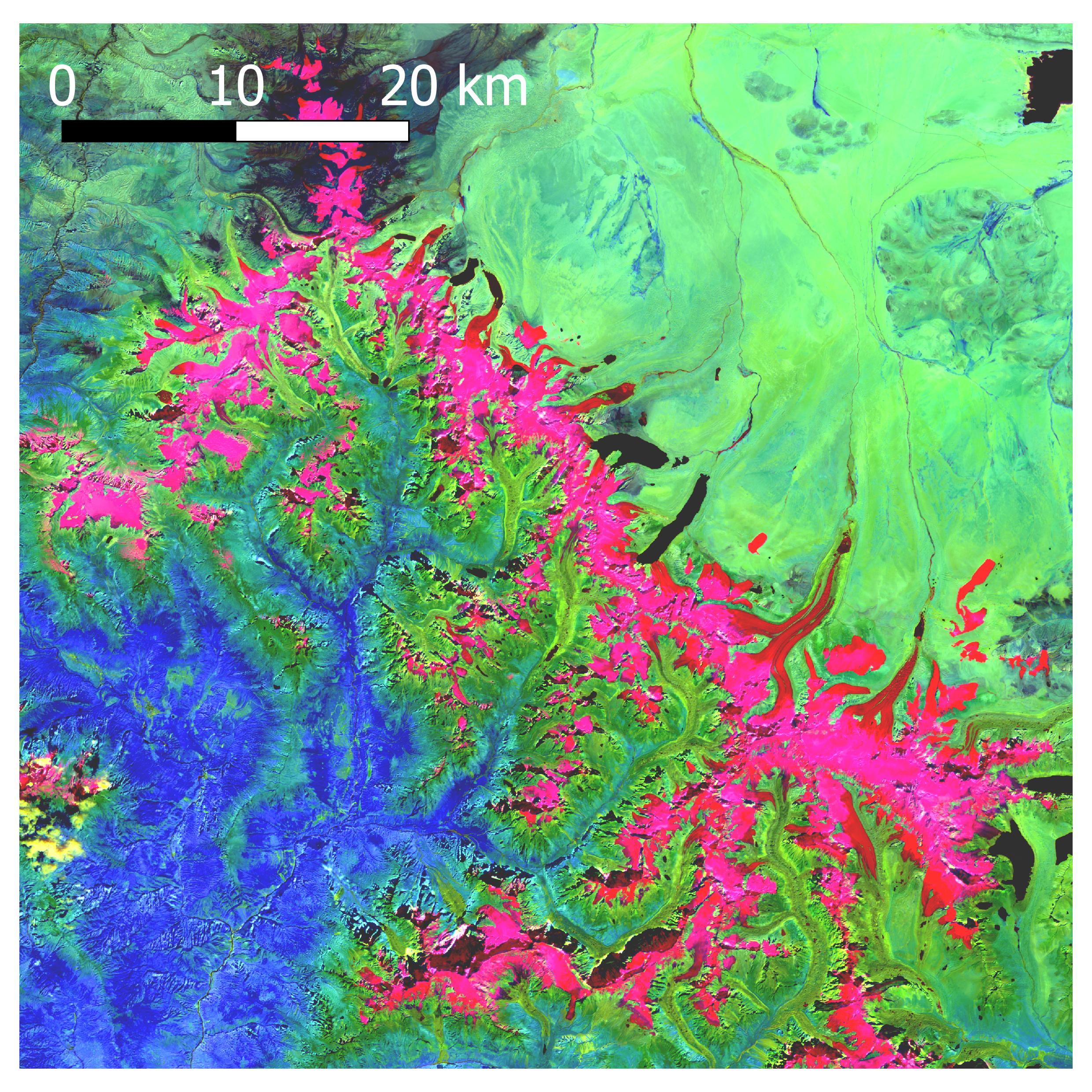}
        \caption{PCA}
        \label{fig:pca}
    \end{subfigure}
    \hfill
    \begin{subfigure}{0.49\columnwidth}
        \includegraphics[width=\linewidth]{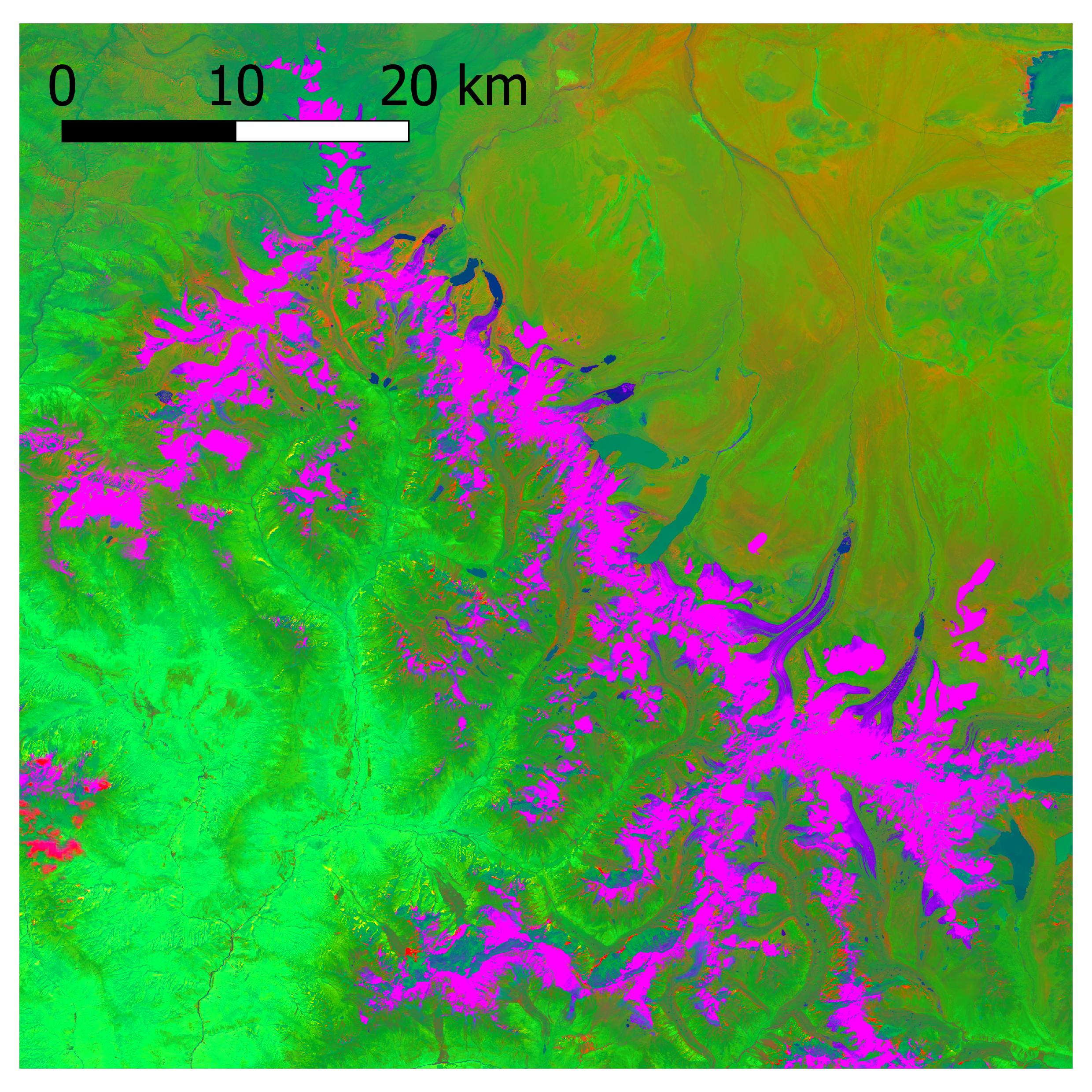}
        \caption{Tasseled Cap (TC)}
        \label{fig:TC}
    \end{subfigure}
    
\caption{LST, PCA, TC, GLCM, and ITS-live velocity used in the analysis.}
\label{fig:features_2x2}
\end{figure}

\textbf{Land Surface Temperature (LST): } The debris layer overlayed on a glacier has a lower temperature until a certain thickness compared to the surrounding debris. Therefore, thermal bands provide valuable information for identifying debris-covered glaciers~\cite{Bolch2007, Ranzi2004, Shukla2010_delineation, Alifu2015}. As Sentinel-2 lacks a thermal band, we used 2 adjacent Landsat-8 scenes acquired on 10 and 15 October 2018 (Fig.~\ref{fig:LST_GL_bound}).

\textbf{Geomorphometric parameters: } there parameters are derived from DEM and improve debris-covered glaciers~\cite{Karimi2012}. We used slope and aspect derived from the Copernicus DEM.
\begin{itemize}
    \item \textit{Slope: }is helpful in debris-covered glacier discrimination, as they have lower slopes on average~\cite{Paul2004}.
    
    \item \textit{Aspect:} representing the direction of slope face, when used along with slope and elevation, enhances the topographic features. 

 \end{itemize}

\textbf{Texture: }is a spatial parameter based on spectral data and helps delineate debris-covered glaciers~\cite{Racoviteanu2012}. Texture features are calculated by the statistical method of the grey-scale co-occurrence matrix (GLCM)~\cite{Lu2021}. It demonstrates how often pairs of grey levels occur at a specific distance. GLCM includes several parameters, such as contrast and homogeneity. In this study, we used dissimilarity, as it highlights heterogeneity. The extracted feature is illustrated in ~\ref{fig2:glcm}.

\textbf{Glacier velocity:} Using velocity data helps discriminate from the stable surrounding lands. For this purpose, we used the Inter-mission Time Series of Land Ice Velocity and Elevation (ITS\_LIVE) ~\cite{gardner2019itslive} from (\url{https://its-live.jpl.nasa.gov/}), a global velocity dataset for annual velocity with 120 m resolution, as shown in Fig.~\ref{fig:ITS}.

\textbf{Interferometric SAR (InSAR) coherence:} Due to the movement, glaciers have lower coherence values than the surrounding, assisting glacier detection~\cite{atwood2010using, Frey2012,  Robson2020}. To generate the coherence feature, two Sentinel-1 SAR images were used. One scene was acquired on the same date as the Sentinel-2 image, and another was obtained 24 days earlier. This temporal baseline was selected as a trade-off between maintaining sufficient signal coherence and enhancing decorrelation over moving glacier surfaces. Compared to shorter intervals (e.g., 12 days), the 24-day interval increases sensitivity to glacier motion, thereby improving the contrast between stable and dynamic surfaces. The resulting coherence image is presented in Fig.~\ref{fig:insar_co}.

\textbf{Principal Component Analysis (PCA): } generates a compressed and more meaningful representation of high-dimensional data by transforming the original data to an orthogonal space. It is used in clean-ice and debris-covered glacier mapping~\cite{Yan2014, Bhambri2011}. In this study, we used 3 PCA features of Sentinel-2. Fig.~\ref{fig:pca} illustrates the RGB visualization of the extracted PCA features.

\textbf{Tasseled Cap (TC) indices:} are linear transformations of optical data to three different features called Tasseled Cap Brightness, Tasseled Cap Greenness, and Tasseled Cap Wetness. Tasseled Cap Brightness represents overall reflectance, Tasseled Cap Greenness is sensitive to vegetation, and Tasseled Cap Wetness is sensitive to moisture content. These indices are informative in glacier mapping, as they are perceived in terms of physical characteristics of the land surface~\cite{Fraser2012}. They help delineate debris-covered glaciers~\cite{Roberts-Pierel2022}. We used all three indices produced by Sentinel-2. Fig.~\ref{fig:TC}shows the RGB color composite of the aforementioned indices.

\begin{figure}[!t]
\centering
\begin{tikzpicture}
    \node[anchor=south west, inner sep=0] (main) at (0,0)
        {\includegraphics[width=1\columnwidth]{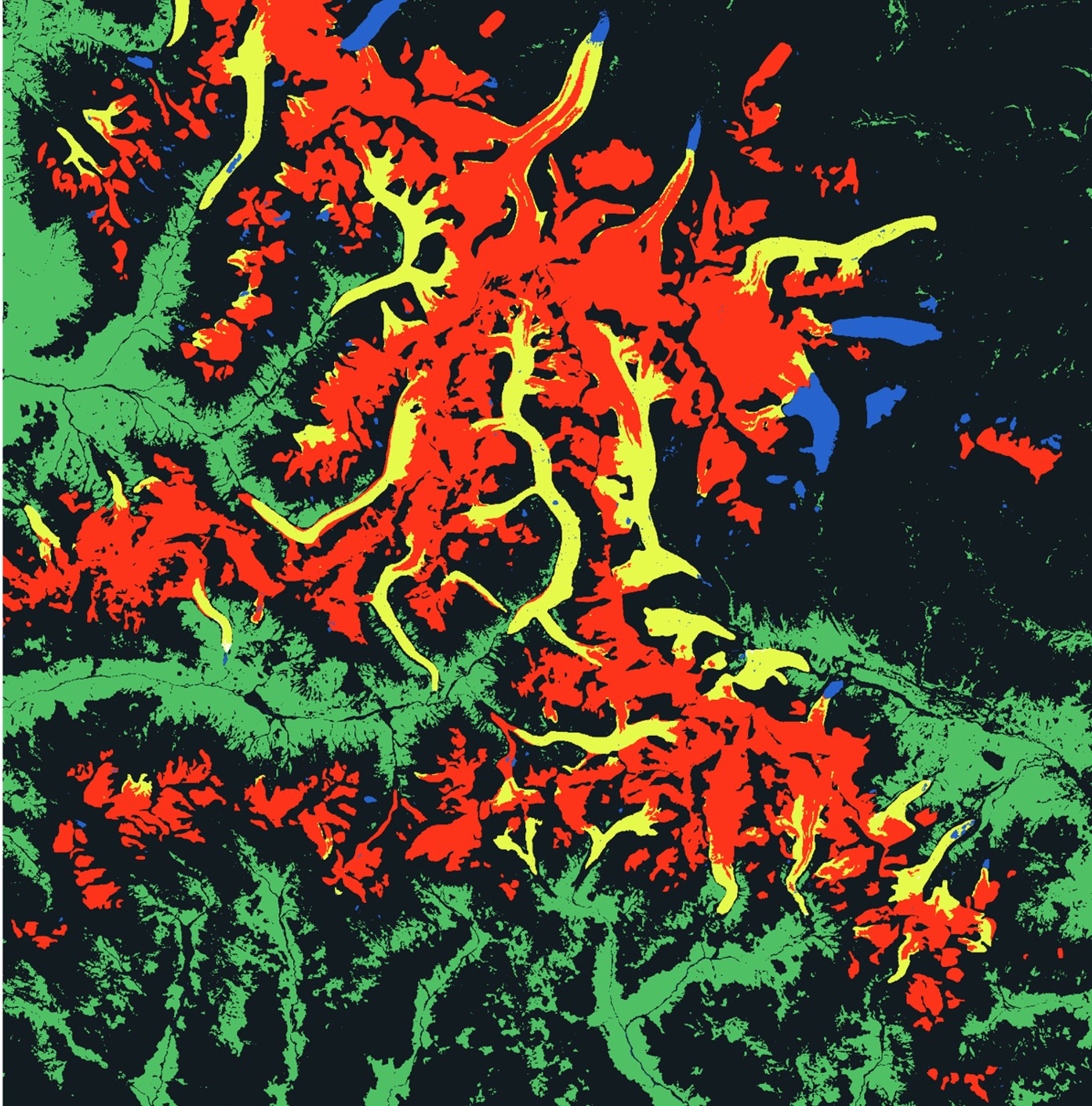}};
    \begin{scope}[x={(main.south east)}, y={(main.north west)}]
        \node[anchor=south west, inner sep=0] at (0.74, 0.86)
            {\includegraphics[width=0.25\columnwidth]
            {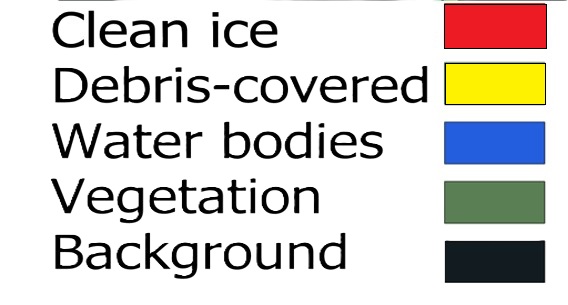}};
        
    \end{scope}
\end{tikzpicture}
\caption{The Reference Labels}
\label{fig:ground_truth}
\end{figure}

\textbf{The Reference Labels:} The same as any supervised semantic segmentation approach, in addition to the images, corresponding label data is required for model training. In this study, five classes are considered, including clean-ice glaciers, debris-covered glaciers, water bodies, vegetation, and background, as is illustrated in Fig~\ref{fig:ground_truth}. The label masks for each class were created as follows:

\begin{itemize}
    
    \item \textit{Debris-covered glaciers parts: }The outline provided by~\cite{Herreid2020}. This dataset was selected due to its global consistency and the use of carefully validated satellite-based mapping, making it suitable for representing the spatial distribution of debris-covered glaciers.
    
    \item \textit{Clean-ice glaciers: } The Randolph Glacier Inventory (RGI) version 7.0 (IACS)~\cite{RGI7.0_2023}, the most recent and further improved version, is used as the primary source of glacier outlines. It is a globally consistent dataset compiled from multiple sources that has been set as close as possible to the year 2000. In our study area, the glacier outlines are primarily derived from ~\cite{Sakai2019}, which presents an updated version of the Glacier Area Mapping for Discharge from the Asian Mountains (GAMDAM) inventory, which represents the first methodologically consistent glacier dataset for HMA. This dataset improves the original inventory by adjusting steep ice and snow-covered slopes and shaded components, based on multi-temporal Landsat imagery. 
    It should be noted that this introduces a temporal mismatch between the reference glacier outlines and the Sentinel-2 imagery used in this study (2018). However, such discrepancies are common in glacier mapping using DL due to the limited temporal availability of global glacier inventories. Despite this limitation, RGI remains widely adopted in previous studies such as~\cite{Maslov2025} and~\cite{Xiang2025}. 
    
    These outlines include both clean-ice and debris-covered glaciers. To obtain clean-ice glacier outlines, the debris-covered polygons were subtracted from the full glacier outlines.
    
    \item \textit{water bodies: }Created using NDWI, and pixel values higher than 0.17 are considered water. The initial outlines were affected by misclassifications in areas such as turbid glacial lakes and shadowed regions. Therefore, manual refinement, such as exploring the whole region to ensure all lakes are correctly delineated, hole filling, and removing misclassified pixels, was applied to correct these errors and improve the accuracy of the water masks. 

    \item \textit{Vegetation: }Created using NDVI, and pixel values higher than 0.3 are considered vegetation.

    \item \textit{Background: }None of the above classes.
\end{itemize}

The final step involved manual refinement of the generated labels. Since the labels were produced from multi-source datasets and a temporal mismatch exists between the reference glacier outlines and the imagery, visually evident inconsistencies between the reference outlines and the imagery were manually corrected, particularly along glacier margins, debris-covered tongues, and lake boundaries.

\subsection{CryoNet: the proposed framework}
\label{sec:model}
We propose an encoder-decoder Convolutional Neural Network (CNN) architecture featuring nested skip connections, enhanced with spatial and channel Squeeze and Excitation (scSE) attention blocks for semantic segmentation. The encoder is based on ResNet101~\cite{he2016deep}, which effectively extracts hierarchical contextual features. We hereafter refer to the proposed architecture as CryoNet.

\begin{figure*}[!t]
\centering
\begin{tikzpicture}
    \node[anchor=south west, inner sep=0] (main) at (0,0)
        {\includegraphics[width=1\textwidth]{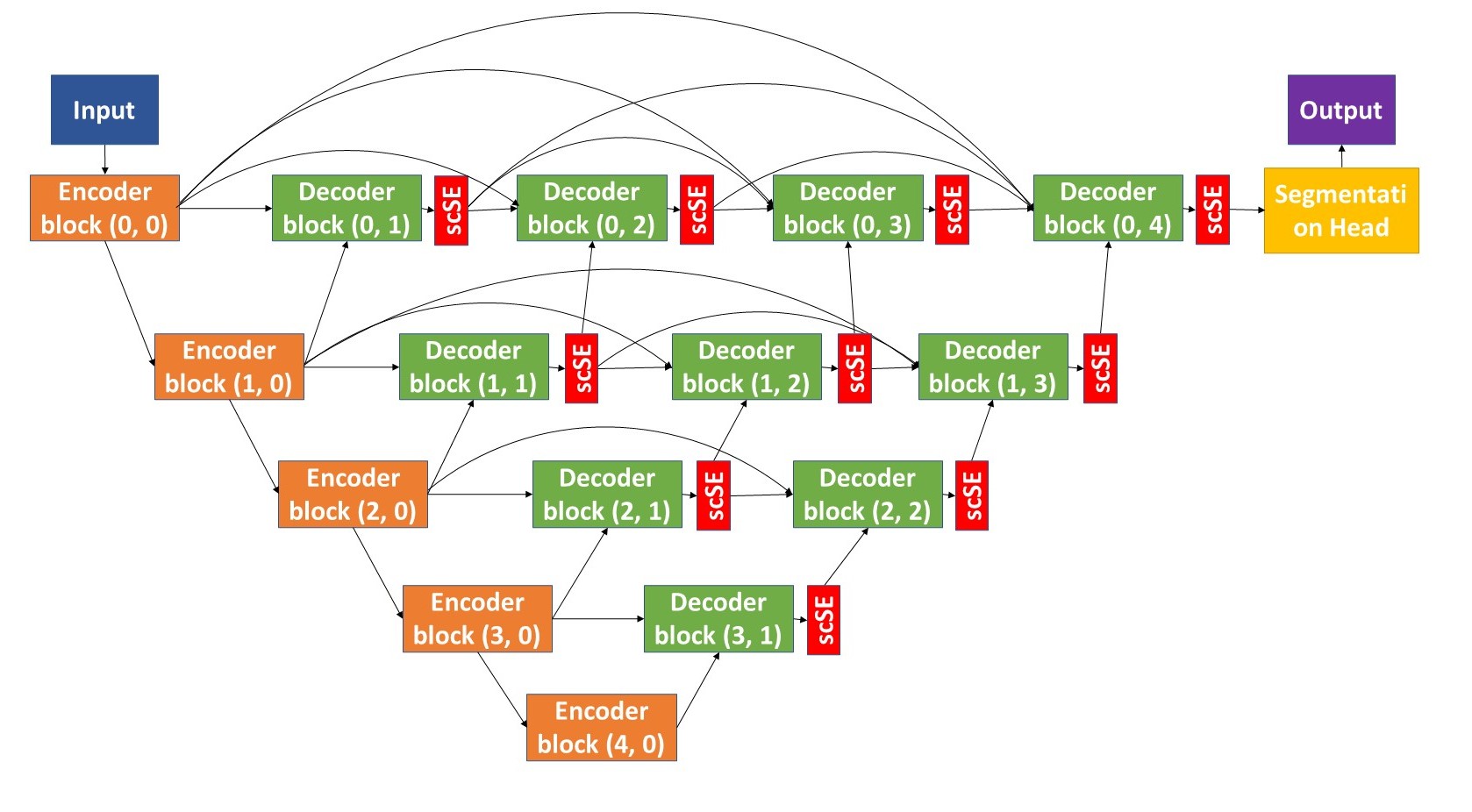}};
        \draw[red,thick] (11,3.9) ellipse [x radius=1.7cm, y radius=0.7cm, rotate=0];
        \draw[->,red,thick] (11.8,3.3) -- (12.7,2.7);
    \begin{scope}[x={(main.south east)}, y={(main.north west)}]
        \node[anchor=south west, inner sep=0] at (0.70, 0.00)
            {\includegraphics[width=0.2\textwidth]
            {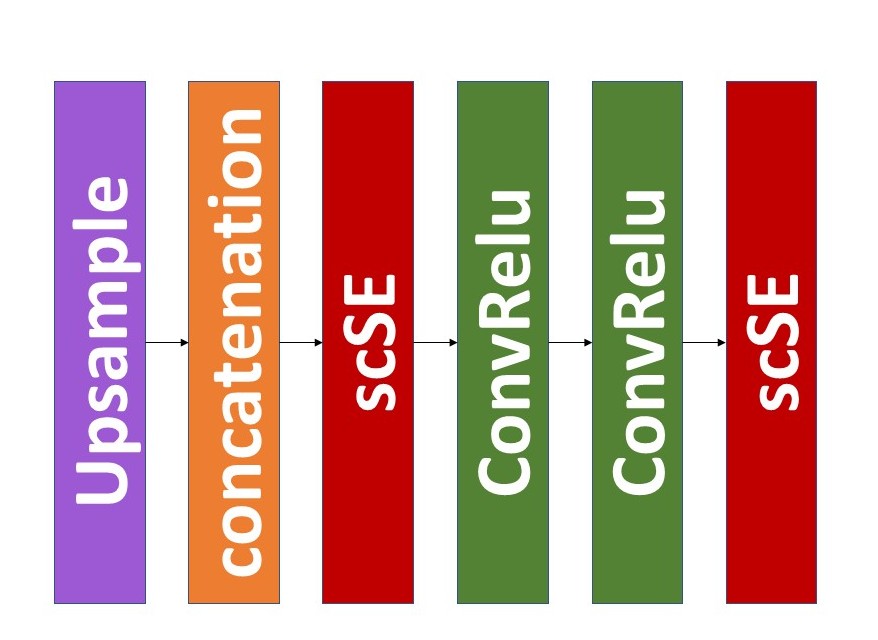}};
        
    \end{scope}
\end{tikzpicture}

\caption{CryoNet architecture}
\label{fig:model}
\end{figure*}

Encoder-decoder CNNs have demonstrated exceptional performance in semantic segmentation tasks. Among these, U-Net and its variations are particularly popular, as their skip connections allow efficient feature propagation between the encoder and decoder while enabling model convergence with relatively fewer training samples~\cite{Ronneberger2015}. The nested skip pathway mechanism extends this concept by embedding multiple U-Nets of varying depths and connecting their decoder layers, thereby facilitating dense feature reuse and multi-scale information flow~\cite{Zhou2020}.

To further enhance representational capability, an attention mechanism is integrated into the network to highlight informative features and suppress less relevant ones~\cite{Hu2018}. The scSE block~\cite{Roy2019} combines channel-wise excitation (cSE), which uses global average pooling to capture the overall context of the input feature map, and spatial excitation (sSE), which recalibrates features at the pixel level through 1x1 convolution to preserve fine spatial details. This dual attention mechanism ensures feature refinement along both channel and spatial dimensions, improving the discriminative power of the network for complex glacier surfaces.

The overall framework of CryoNet is schematically illustrated in Fig.\ref{fig:model}. The model takes the  30-layer multi-source feature stack as described in\ref{sec:data} as the input and feeds it into the ResNet101 encoder. The orange blocks represent the main encoder stages. Encoder block (0,0) includes convolutional, batch normalization, ReLU, and max pooling layers. Encoder blocks (1,0) to (4,0) correspond to the four residual stages of ResNet101, containing 3, 4, 23, and 3 bottleneck blocks, respectively. Through successive convolutional and downsampling operations, the encoder extracts hierarchical representations at multiple spatial scales. The green blocks represent the decoder modules, each comprising upsampling, concatenation, convolutional, and ReLU layers. The nested skip pathways densely connect encoder and decoder stages, enabling the reuse and aggregation of intermediate feature maps across resolutions. The red blocks denote scSE attention blocks. Each block, embedded within its corresponding decoder block, refines features by suppressing less informative features and emphasizing the most effective ones before the upsampling stages. This ensures that both global context and local details contribute to the final pixel-wise classification map of the five glacier-related surface classes.

The final decoder output is fed into the segmentation head, which produces a five-channel probability map corresponding to clean ice, debris-covered ice, water, vegetation, and background classes, followed by the final segmentation output map.

This architecture is well-suited to address the spectral and spatial complexity of glacier surfaces, particularly the low contrast between debris-covered ice and surrounding non-glacial terrain. By leveraging multi-scale feature fusion and joint spatial–channel attention, the model enhances class separability and boundary delineation in heterogeneous mountain environments. The training configuration is detailed in the following section.

\subsection{Training and testing strategy}
\label{sec:training}
To train CryoNet, we conducted a series of experiments to ensure stable optimization, class balance, and reproducible performance.

The model contains ~70 M trainable parameters. The network input consisted of 30-band image patches of size 256 × 256 pixels. Patch sizes of 128 and 512 pixels were also tested; however, the best results were achieved with 256×256 pixels. The resulting patch set was randomly split into 80\% as the training dataset and 20\% as the test dataset. The model was trained from scratch, without using any pretrained weights. 

Training was performed for 100 epochs with a batch size of 16. Several optimizers, including Adam and stochastic gradient descent (SGD) with momentum, were initially tested. The Adam optimizer provided the most stable convergence and was therefore adopted in the final experiments. 
The learning rate was set to $3 \times 10^{-4}$ and the weight decay to $1 \times 10^{-4}$ to balance stability and convergence speed while reducing the risk of overfitting. To address class imbalance, class-frequency weighting was applied.

The weighted cross-entropy loss was used as the main objective function. Alternative overlap-based losses, such as Dice and Tversky, were evaluated but ultimately omitted, as the weighted Cross-Entropy achieved the best performance in terms of both accuracy and stability.

\subsection{Evaluation metrics}
\label{sec:evaluate}
The performance of CryoNet was assessed through both quantitative and qualitative evaluations. Quantitative metrics were computed on the test dataset, and qualitative evaluation was performed on an independent, unseen subset of the data.

The metrics used for quantitative evaluation include per-class Intersection over Union (IoU), Precision, Recall, and their corresponding mean values per class. Equation \ref{equ:acc} defines these metrics.

\begin{subequations}
\centering
\begin{align}
    \textrm{IoU} &= \frac{TP}{TP+FP+FN}\\
    \textrm{Precision} &= \frac{TP}{TP+FP}\\
    \textrm{Recall} &= \frac{TP}{TP+FN}\\
    \textrm{Total Accuracy} &= \frac{\textrm{Total TP}}{\textrm{Total Pixels}}
\end{align}
\label{equ:acc}
\end{subequations}

where TP, FP, and FN denote the number of true-positive, false-positive, and false-negative pixels, respectively.

To further validate performance, the results achieved by CryoNet were compared with three state-of-the-art semantic segmentation architectures, including DeepLabV3+~\cite{10.1007/978-3-030-01234-2_49}, SegFormer~\cite{NEURIPS2021_64f1f27b}, and U-Net~\cite{Ronneberger2015}, trained under identical conditions and data preprocessing settings. 

To reconstruct the full-size segmentation maps, overlapping patches were merged using a Hanning-weighted unpatchification approach. The Hanning window, which is a smooth cosine-shaped weighting function, assigns higher weights to the patch center and gradually reduces them toward the edges. This smoothing minimized edge artifacts and ensured seamless transitions between adjacent patches.

\subsection{Channel Importance Analysis}
\label{sec:channel_importance}
To understand how much each input band contributed to the performance of the segmentation model, we conducted a permutation-based channel importance analysis. This approach measures how sensitive the model is to each band by assessing how much the segmentation accuracy decreases when the spatial structure of that band is intentionally disrupted.

First, the trained model was evaluated on the original, unmodified test set to obtain baseline values for mean Intersection over Union (mIoU) and accuracy (Acc). Then, for every band, we created a modified version of the test data in which the values of that band were randomly shuffled across samples. This preserves the overall distribution of the band but destroys the meaningful spatial patterns that the model relies on. The model was run again on each of these perturbed datasets. The importance of a band was determined by how much the IoU and accuracy dropped compared to the baseline.

Because different glacier surface types rely on different physical and spectral characteristics, the permutation test was performed separately for clean-ice and debris-covered glaciers as well. This allowed us to identify which input bands were most relevant for each glacier type.

\subsection{Transferability}
\label{sec:transferability}  

\begin{figure*}[t]
\centering
\begin{tikzpicture}
    \node[anchor=south west, inner sep=0] (main) at (0,0)
        {\includegraphics[width=1\textwidth]{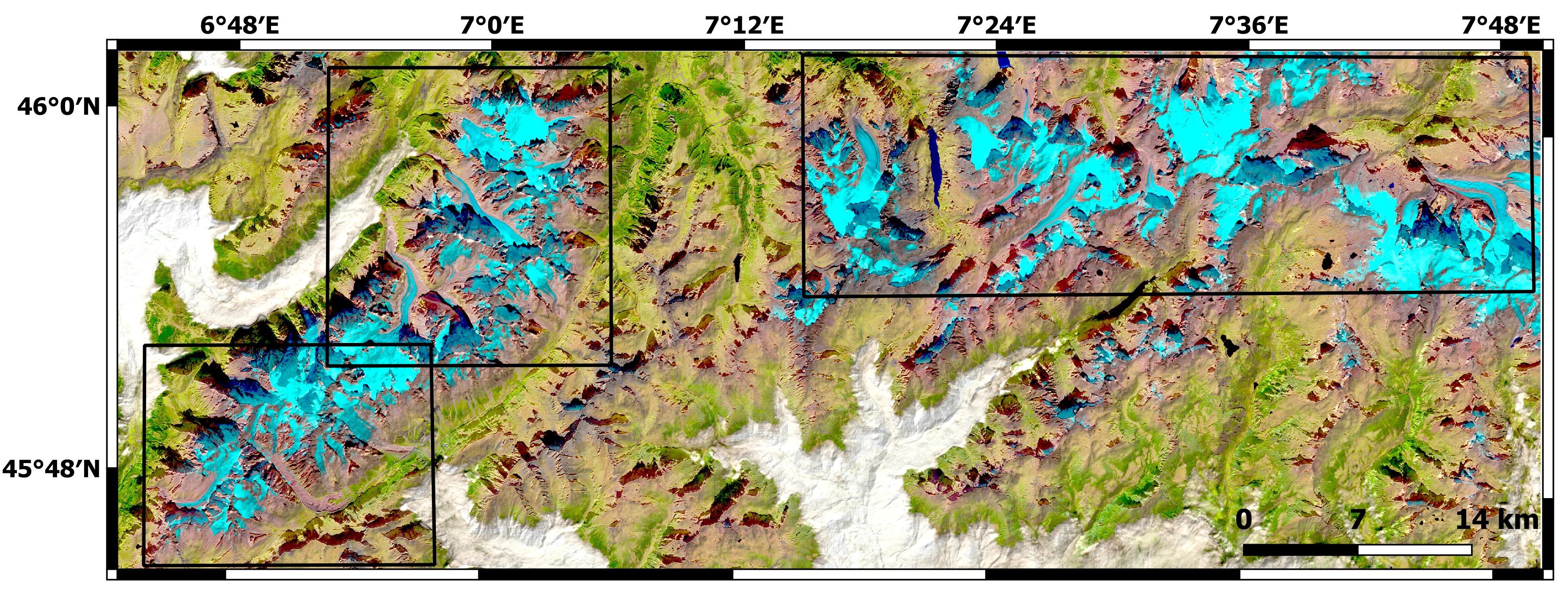}};
    \begin{scope}[x={(main.south east)}, y={(main.north west)}]
        \node[anchor=south west, inner sep=0] at (0.4, -0.06)
            {\includegraphics[width=0.45\columnwidth]{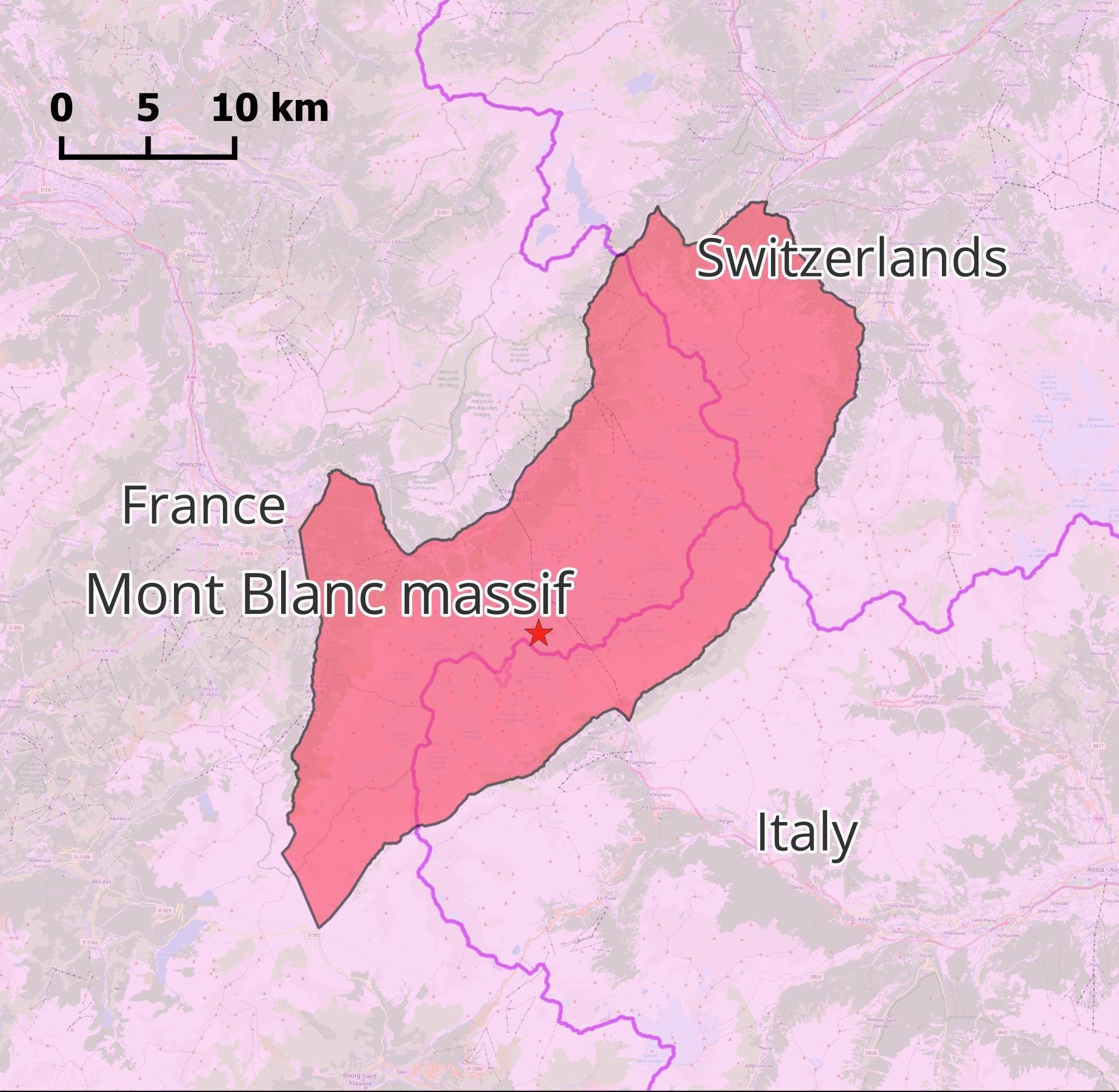}};
    \end{scope}
\end{tikzpicture}
\caption{Mont Blanc Massif Region. The black boxes show the areas of interest used in this study.}
\label{fig:montblanc}
\end{figure*}

To assess the model’s transferability, we applied the trained model to an independent region. The selected study area is located in the Mont Blanc massif in the Alps, near the border of France, Italy, and Switzerland (see Fig.~\ref{fig:montblanc}). The dataset for this region was generated following the same procedure as the training data. The corresponding Sentinel-2 image was acquired on 22 October 2018. It was selected due to its minimal cloud coverage and overall favorable surface conditions. In this scene, seasonal snow cover remains limited. Additionally, this date is consistent with the data used to train the model. The pre-trained model was subsequently fine-tuned for a limited number of iterations to better adapt to the characteristics of the new region. 

\section{Results}\label{sec:results}

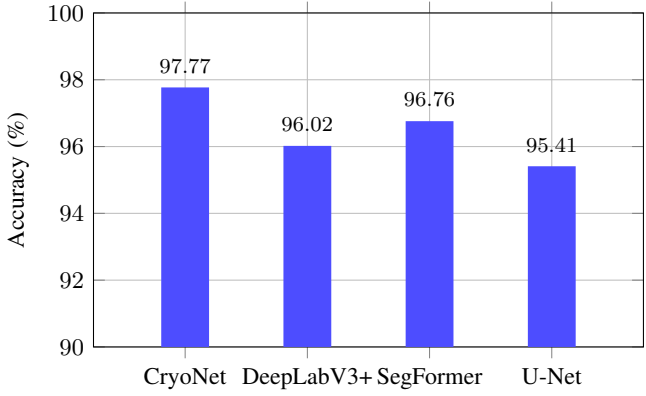
\begin{figure}[!ht]
\centering
\begin{tikzpicture}
\begin{axis}[
    ybar,
    ymin=90, ymax=100,
    bar width=18pt,
    enlarge x limits=0.25,
    ylabel={Accuracy (\%)},
    symbolic x coords={CryoNet, DeepLabV3+, SegFormer, U-Net},
    xtick=data,
    nodes near coords,
    nodes near coords align={vertical},
    every node near coord/.append style={font=\footnotesize, yshift=2pt},
    width=\columnwidth,
    height=6cm,
    grid=major,
    ymajorgrids=true,
    tick label style={font=\small},
    ylabel style={font=\small},
    bar shift=0pt,
]
\addplot[fill=blue!70, draw=none] coordinates {
    (CryoNet,97.77)
    (DeepLabV3+,96.02)
    (SegFormer,96.76)
    (U-Net,95.41)
};
\end{axis}
\end{tikzpicture}
\caption{Comparison of overall accuracy between different models.}
\label{chart:overal_acc}
\end{figure}

The quantitative and qualitative evaluations demonstrate that CryoNet outperformed the three benchmark architectures trained under the same conditions. As Fig.\ref{chart:overal_acc} shows, CryoNet achieved the highest overall accuracy with 97.77\%, outperforming DeepLabV3+(96.02\%), SegFormer(96.76\%), and U-Net(95.41\%). 

Moreover, class-wise inspection emphasizes the performance of CryoNet. Tables\ref{tab:iou}, \ref{tab:recall}, and \ref{tab:precision} summarize per-class and mean results for IoU, recall, and precision, respectively, for the Poiqu Basin study area. Our proposed model is highlighted in bold, and the best results in each column are also shown in bold. CryoNet obtained the best mean IoU (90.52\%), mean Recall (98.08\%), and mean Precision (92.26\%) among all models. A significant improvement was observed for debris-covered glaciers, where CryoNet reached an IoU of 90.46\%.

\begin{table*}[!t]
    \centering
    \resizebox{=\textwidth}{!}{
    \begin{tabular}{lccccccccccr}
         \toprule
         Model & & IoU & & & & Mean IoU\\
               & clean-ice glaciers & Debris-covered glaciers & Water bodies &  Vegetation & Background &  \\
         \midrule
         \textbf{CryoNet}                      & \textbf{94.01} & \textbf{90.46} & \textbf{96.5} & \textbf{74.40} & \textbf{97.22} & \textbf{90.52}\\
         DeepLabV3+ \cite{10.1007/978-3-030-01234-2_49}           & 89.87 & 82.32 & 91.82 & 61.49 & 95.15 & 84.13 \\
         SegFormer \cite{NEURIPS2021_64f1f27b}           & 91.79 & 86.30 & 92.18 & 63.07 & 96.01 & 85.87 \\
         U-Net \cite{Ronneberger2015}           & 87.25 & 79.41 & 84.44 & 72.02 & 94.52 & 83.53 \\
         \bottomrule\\
    \end{tabular}}
    \caption{IoU for the Poiqu Basin study area}
    \label{tab:iou}
\end{table*}

\begin{table*}[!ht]
    \centering
    \resizebox{=\textwidth}{!}{
    \begin{tabular}{lccccccccccr}
         \toprule
         Model & & Recall & & & & Mean Recall\\
               & clean-ice glaciers & Debris-covered glaciers & Water bodies &  Vegetation & Background &  \\
         \midrule
         \textbf{CryoNet}          & \textbf{97.30} & \textbf{95.79} & 99.61 & \textbf{99.64} & \textbf{98.09} & \textbf{98.08}\\
         DeepLabV3+            & 95.32 & 92.84 & 99.48 & 97.64 & 96.53 & 96.36 \\
         SegFormer            & 95.40 & 95.22 & 99.36 & 97.44 & 97.39 & 96.96 \\
         U-Net            & 91.96 & 92.44 & \textbf{99.65} & 97.80 & 96.92 & 95.75 \\
         \bottomrule\\
    \end{tabular}}
    \caption{Recall for the Poiqu Basin study area}
    \label{tab:recall}
\end{table*}

\begin{table*}[!t]
    \centering
    \resizebox{=\textwidth}{!}{
    \begin{tabular}{lccccccccccr}
         \toprule
         Model & & Precision & & & & Mean Precision\\
               & clean-ice glaciers & Debris-covered glaciers & Water bodies &  Vegetation & Background &  \\
         \midrule
         \textbf{CryoNet}          & \textbf{96.53} & \textbf{94.21} & \textbf{96.87} & \textbf{74.60} & \textbf{99.10} & \textbf{92.26}\\
         DeepLabV3+            & 94.02 & 87.90 & 92.27 & 64.42 & 98.52 & 87.02\\
         SegFormer            & 96.03 & 90.22 & 92.73 & 64.13 & 98.54 & 88.33 \\
         U-Net            & 94.46 & 84.93 & 84.69 & 73.20 & 97.45 & 86.95 \\
         \bottomrule\\
    \end{tabular}}
    \caption{Precision for the Poiqu Basin study area}
    \label{tab:precision}
\end{table*}

A higher IoU value indicates that the predicted outlines overlap more accurately with the ground-truth masks, meaning CryoNet generates glacier boundaries that are spatially closer to expert delineations. In particular, the IoU achieved for debris-covered glaciers demonstrates that the model can correctly identify most of the pixels belonging to these complex regions while minimizing false inclusions of surrounding terrain.

High recall signifies that the model successfully captures most correct pixels, while very few areas are missed. This is critical for hazard and volume-change assessment, as omission errors in glacier inventories would lead to underestimation of glacier extents.

High precision indicates strong reliability in boundary delineation. Meanwhile, strong precision shows that most predicted pixels truly belong to the target class, confirming that CryoNet effectively suppresses false detections in spectrally similar non-glacial areas.

\begin{figure*}[!t]
\centering
\begin{subfigure}{\columnwidth}
    \includegraphics[width=\linewidth]{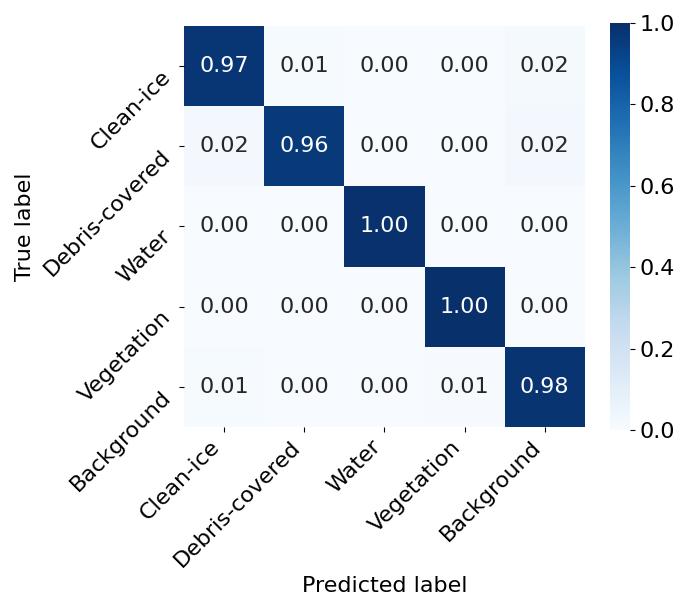}
    \caption{CryoNet}
    \label{fig:Cryo_conf_mat}
\end{subfigure}
\hfill
\begin{subfigure}{\columnwidth}
    \includegraphics[width=\linewidth]{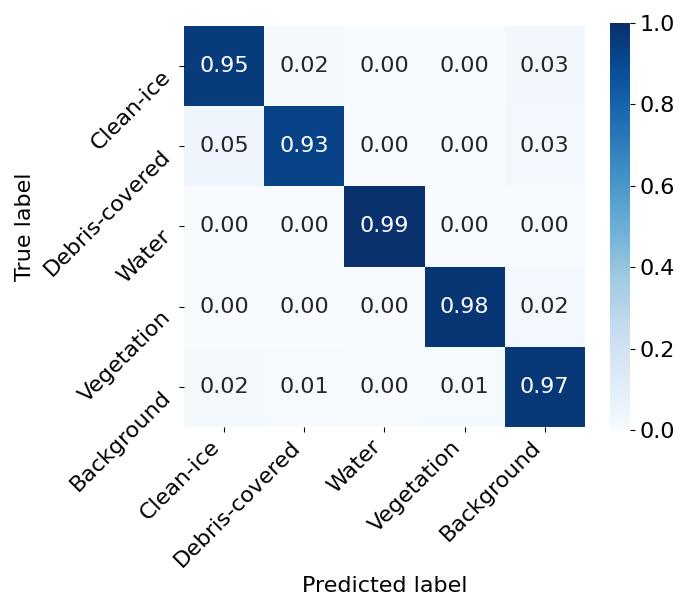}
    \caption{DeepLabV3+}
    \label{fig:deep_conf_mat}
\end{subfigure}    
\vspace{2mm}
\begin{subfigure}{\columnwidth}
    \includegraphics[width=\linewidth]{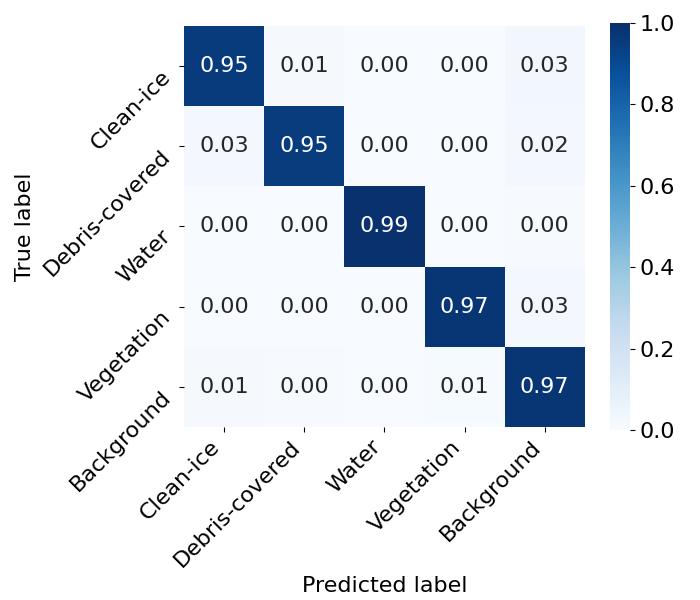}
    \caption{SegFormer}
    \label{fig:segformer_conf_mat}
\end{subfigure}
\hfill
\begin{subfigure}{\columnwidth}
    \includegraphics[width=\linewidth]{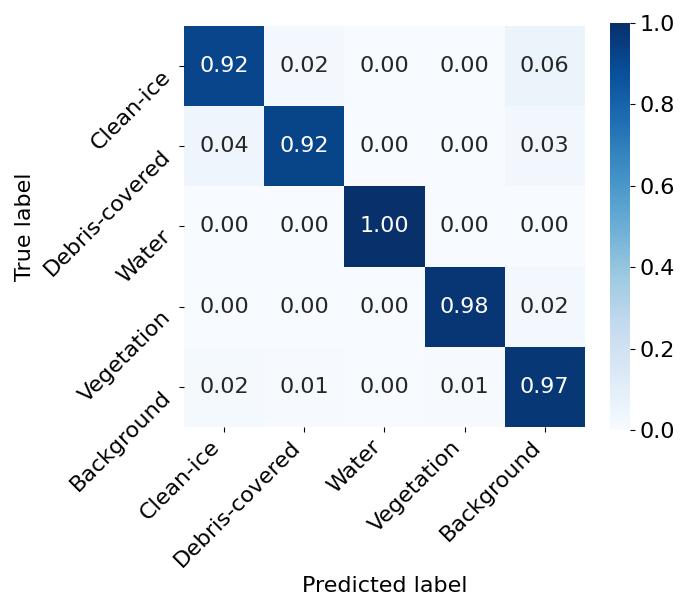}
    \caption{U-Net}
    \label{fig:unet_conf_mat}
\end{subfigure}
\caption{Confusion matrices}
\label{fig:conf_mat}
\end{figure*}

The confusion matrices presented in Fig.\ref{fig:conf_mat} provide a detailed overview of the pixel-wise classification behavior of CryoNet compared to the three models. Across all classes, CryoNet demonstrates the lowest off-diagonal values, indicating fewer misclassifications. Most confusion in DeepLabV3+, SegFormer, and U-Net occurs between debris-covered glaciers and the surrounding background, reflecting the spectral similarity of these surfaces. In contrast, CryoNet correctly identifies the majority of debris-covered pixels and suppresses false assignments to non-glacial terrain, which is consistent with its high IoU for this class. 

\begin{table*}[!t]
    \centering
    \resizebox{=\textwidth}{!}{
    
    \begin{tabular}{l|cccccccccc}
    \shortstack{Test image\\(True Colour)}&\includegraphics[width=0.16\columnwidth]{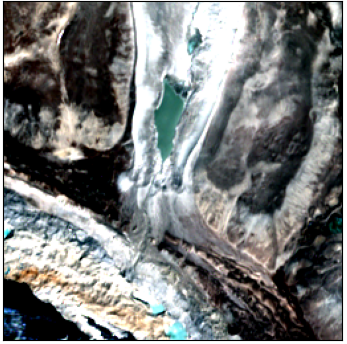} & 
    \includegraphics[width=0.16\columnwidth]{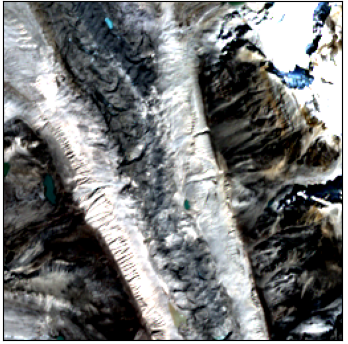} &
    \includegraphics[width=0.16\columnwidth]{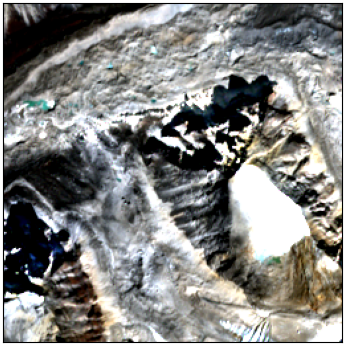} & 
    \includegraphics[width=0.16\columnwidth]{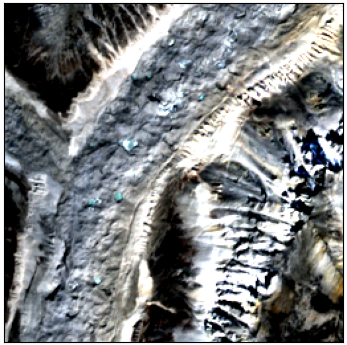} & 
    \includegraphics[width=0.16\columnwidth]{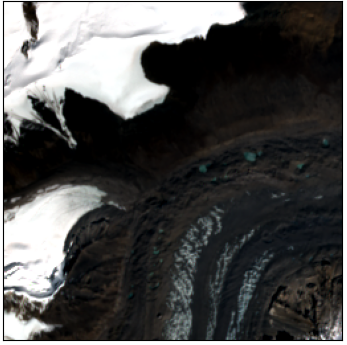} & 
    \includegraphics[width=0.16\columnwidth]{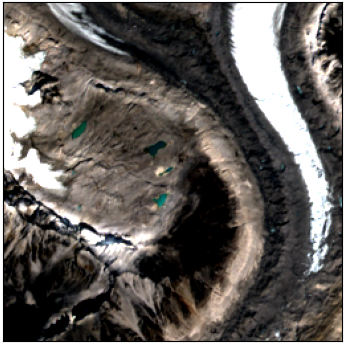} & 
    \includegraphics[width=0.16\columnwidth]{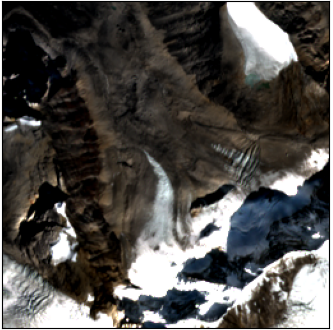} & 
    \includegraphics[width=0.16\columnwidth]{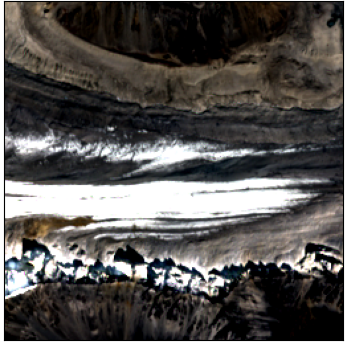} \\

    \shortstack{Test image\\(False colour)} &\includegraphics[width=0.16\columnwidth]{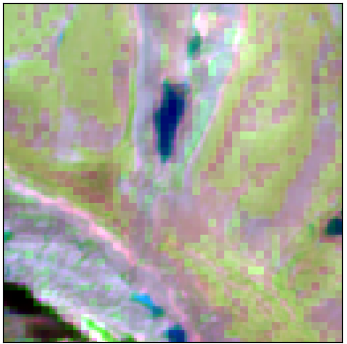} & 
    \includegraphics[width=0.16\columnwidth]{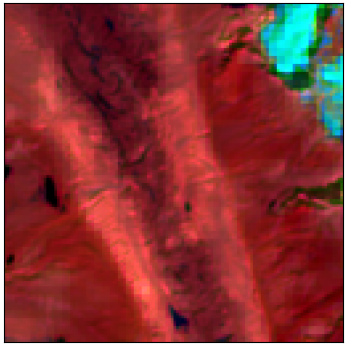} &
    \includegraphics[width=0.16\columnwidth]{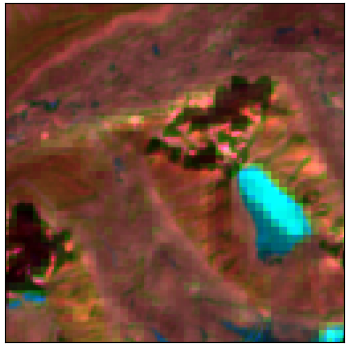} & 
    \includegraphics[width=0.16\columnwidth]{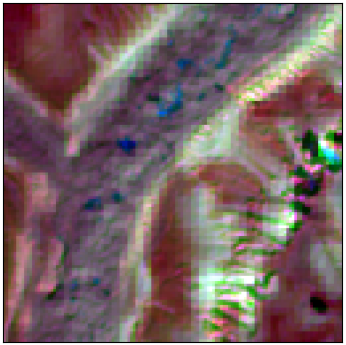} & 
    \includegraphics[width=0.16\columnwidth]{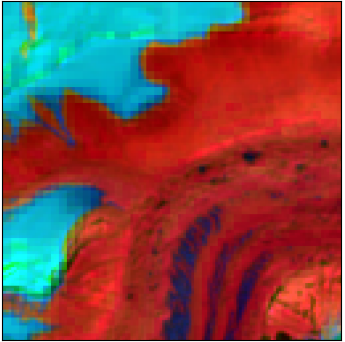} & 
    \includegraphics[width=0.16\columnwidth]{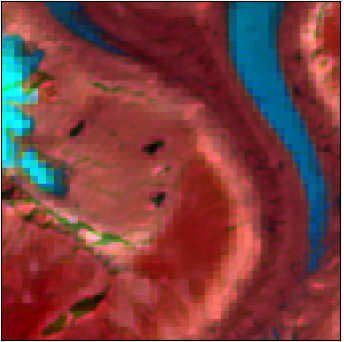} & 
    \includegraphics[width=0.16\columnwidth]{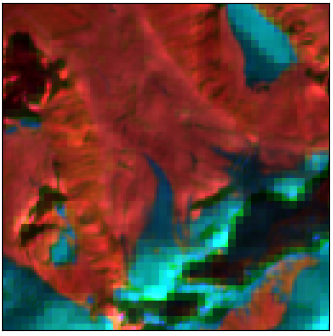} & 
    \includegraphics[width=0.16\columnwidth]{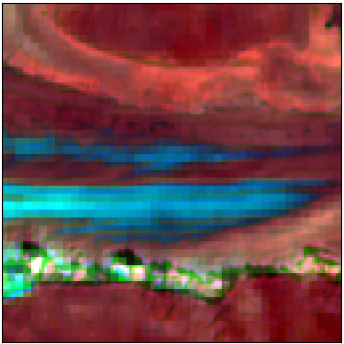}\\
    
    The reference labels &\includegraphics[width=0.16\columnwidth]{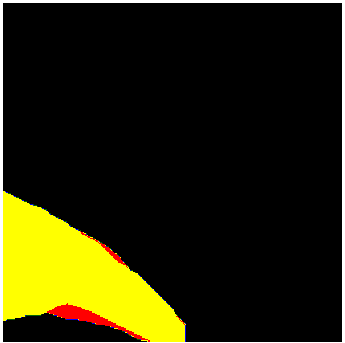} &
    \includegraphics[width=0.16\columnwidth]{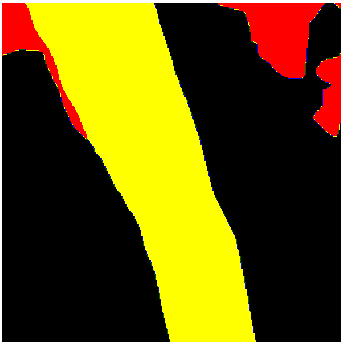} &
    \includegraphics[width=0.16\columnwidth]{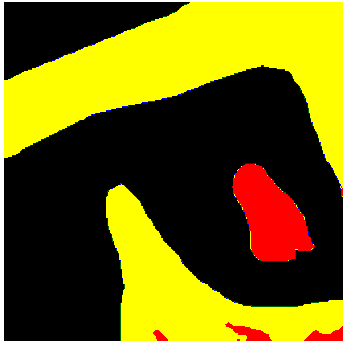} & 
    \includegraphics[width=0.16\columnwidth]{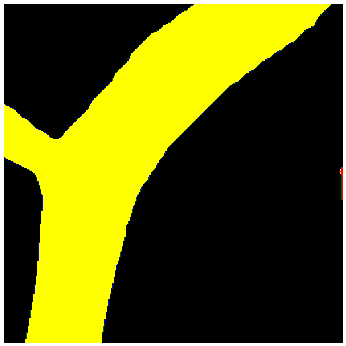} & 
    \includegraphics[width=0.16\columnwidth]{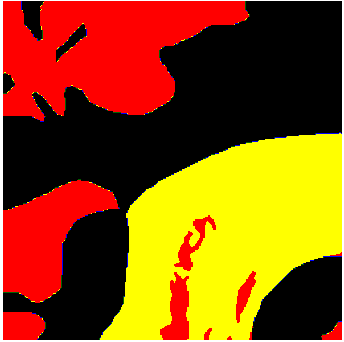} & 
    \includegraphics[width=0.16\columnwidth]{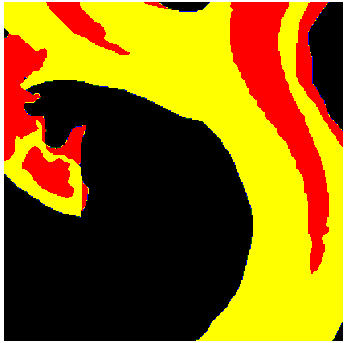} & 
    \includegraphics[width=0.16\columnwidth]{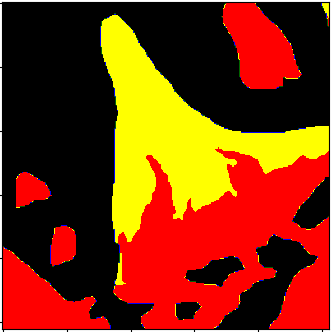} & 
    \includegraphics[width=0.16\columnwidth]{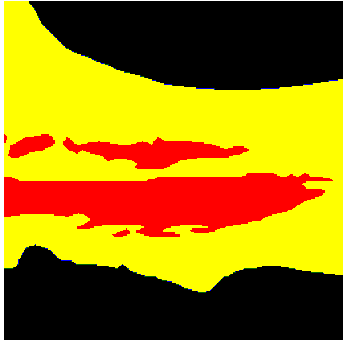} \\

    Prediction &\includegraphics[width=0.16\columnwidth]{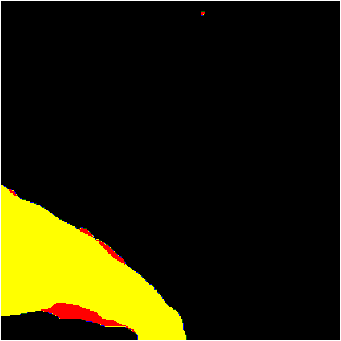} &
    \includegraphics[width=0.16\columnwidth]{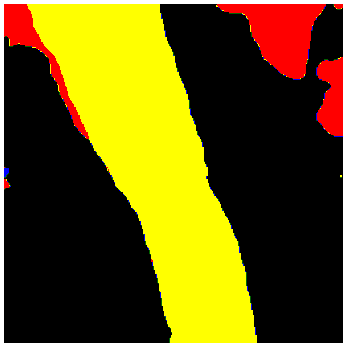} &
    \includegraphics[width=0.16\columnwidth]{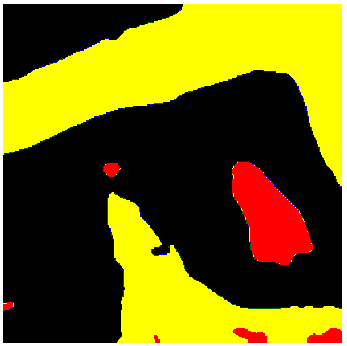} & 
    \includegraphics[width=0.16\columnwidth]{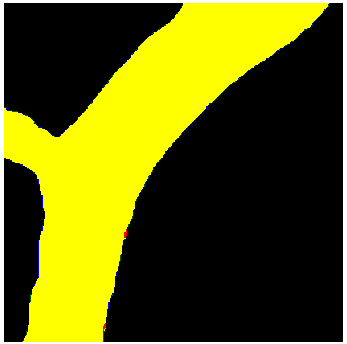} & 
    \includegraphics[width=0.16\columnwidth]{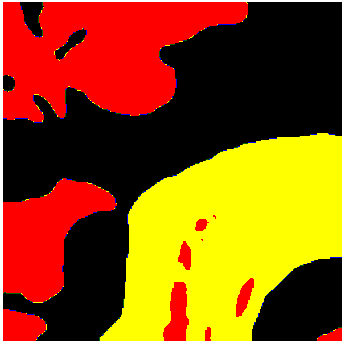} & 
    \includegraphics[width=0.16\columnwidth]{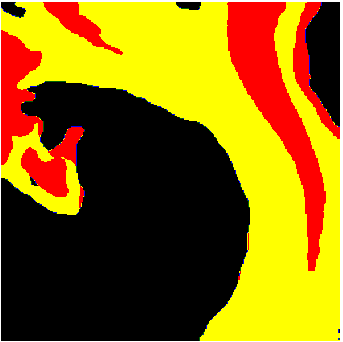} & 
    \includegraphics[width=0.16\columnwidth]{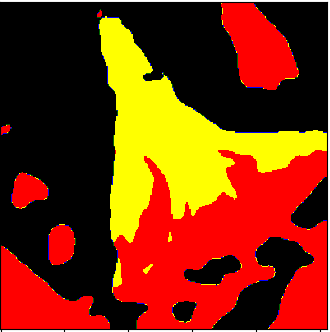} & 
    \includegraphics[width=0.16\columnwidth]{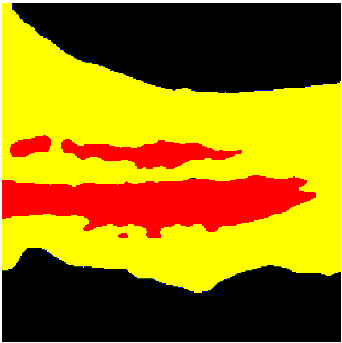} \\

    \end{tabular}}
    \caption{Example results of CryoNet on the test data.}
    \label{tab:qualitative_test_predict}
\end{table*}

\begin{figure*}[!t]
    \centering
    
    \begin{subfigure}{0.32\textwidth}
        \includegraphics[width=\columnwidth]{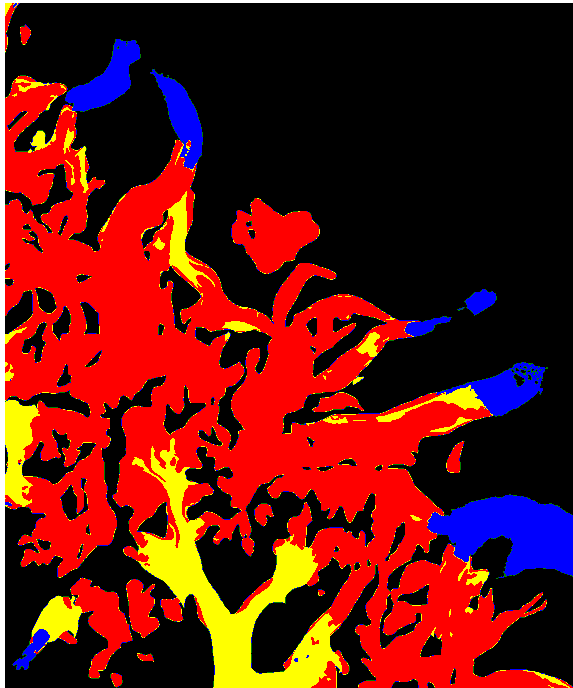}
        \caption{The label mask}
        \label{fig:GT_subset}
    \end{subfigure}
   \hfill
    \begin{subfigure}{0.32\textwidth}
        \includegraphics[width=\columnwidth]{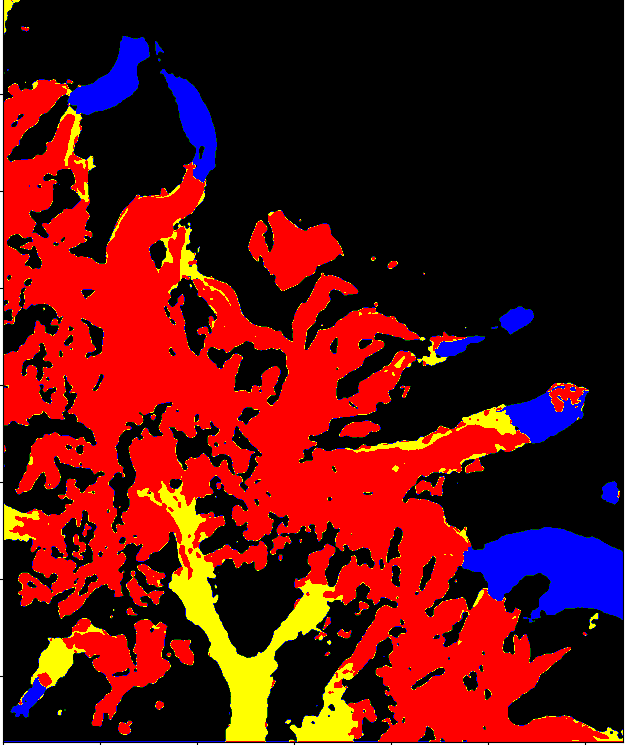}
        \caption{CryoNet}
        \label{fig:result_cryonet_subset}
    \end{subfigure} 
    \hfill
    \begin{subfigure}{0.32\textwidth}
        \begin{tikzpicture}
        \node[anchor=south west, inner sep=0] (main) at (0,0)
            {\includegraphics[width=1\columnwidth]{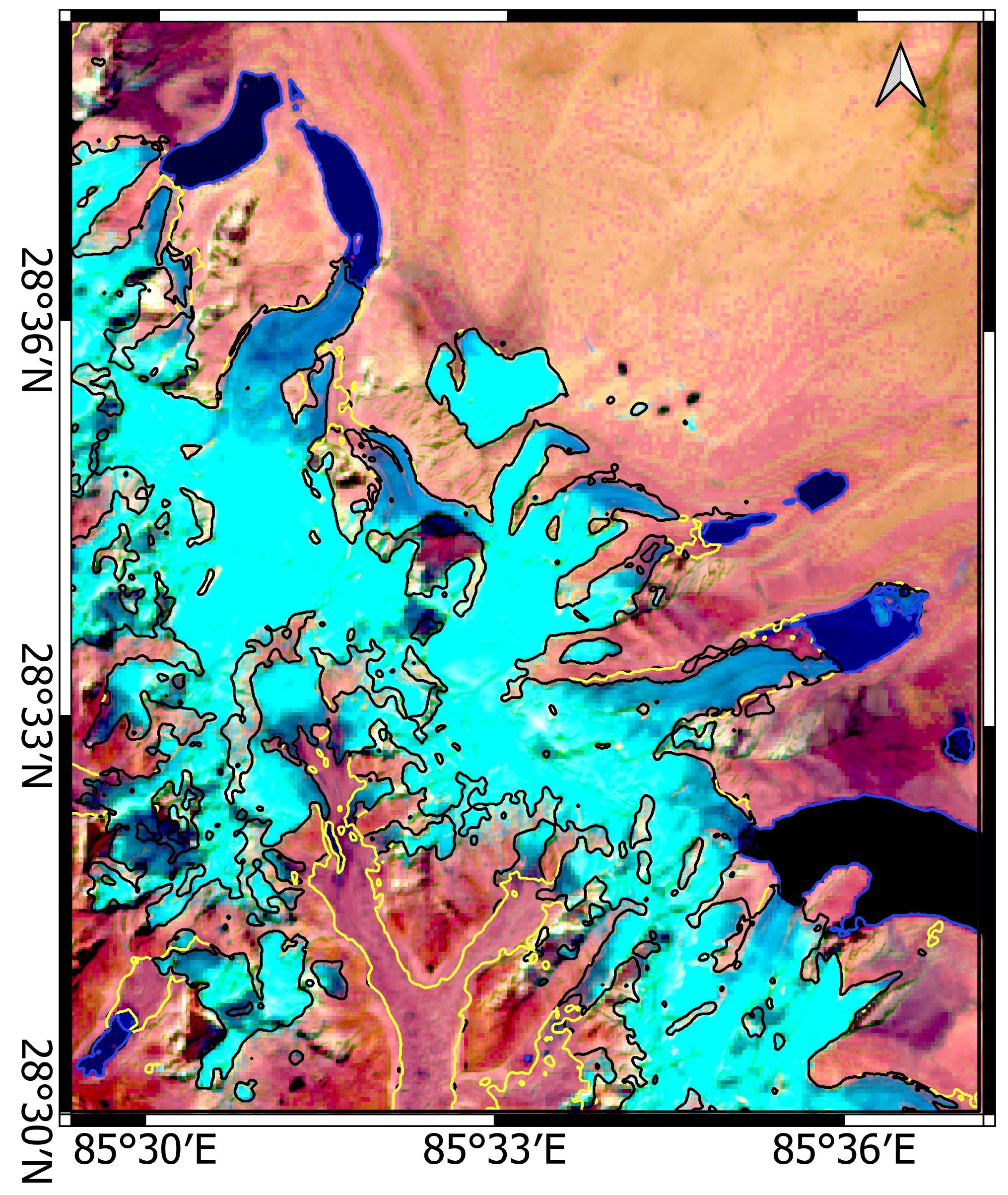}};
            \begin{scope}[x={(main.south east)}, y={(main.north west)}]
                \node[anchor=south west, inner sep=0] at (0.40, 0.84)
                    {\includegraphics[width=0.45\columnwidth]{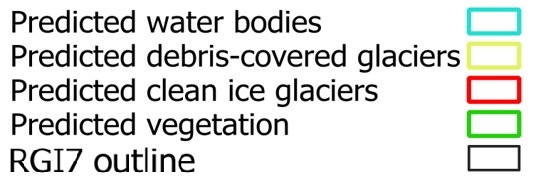}};
            \end{scope}
        \end{tikzpicture}
        \caption{Predictions on Sentinel-2 image}
        \label{fig:result_sen2}
    \end{subfigure} 
\caption{Prediction results on an unseen subset: (a) the corresponding ground truth, (b) Predictions of CryoNet, (c) Predictions on Sentinel-2 image}
\label{fig:result_imgs}
\end{figure*}

The qualitative patch-level comparisons of test data in Table\ref{tab:qualitative_test_predict} further support these findings. The prediction of CryoNet is well aligned with the ground truth. The model preserves the continuity, curvature, and sharp transitions of glacier boundaries even in complex patches. CryoNet accurately captures even the narrow, elongated debris-covered areas, which are typically medial and side moraines and are often challenging to delineate. Moreover, it is noticeable that although the ground-truth boundaries are not perfectly defined, the model produces smoother and more realistic glacier outlines that fit the used data much better.  
Fig.\ref{fig:result_imgs} illustrates the prediction from CryoNet on unseen data without any post-processing versus the corresponding labels as described in \ref{sec:data}. Fig.\ref{fig:result_sen2} shows the same subset overlaid on the Sentinel 2 image in false color composite (RGB: B11, B9, B4). Qualitative inspection emphasizes the performance of CryoNet. The predicted segmentation map aligns closely with the ground-truth labels, confirming that the model generalizes well beyond the training data. Fine topographic and textural details, such as glacier tongues, narrow ice flows, and boundary curvature, are preserved.

Table~\ref{tab:alps_acc_table} presents the quantitative evaluation results for the Mont Blanc Massif, used as an independent test region to assess the transferability of the trained model. The results demonstrate that CryoNet maintains strong performance, achieving an overall accuracy of 96.32\%. These results confirm that CryoNet preserves a high level of segmentation accuracy when applied to a geographically distinct region, demonstrating promising transferability, particularly when minimal fine-tuning is applied to account for regional differences.

\begin{table}[!t]
    \centering
    \resizebox{\columnwidth}{!}{
    \begin{tabular}{lccc}
         \toprule
               & IoU & Precision & Recall  \\
         \midrule
         Clean-ice glaciers      & 93.37 & 96.85 & 96.30 \\
         Debris-covered glaciers & 84.15 & 88.05 & 94.99 \\
         Water bodies            & 95.55 & 95.69 & 99.85 \\
         Vegetation              & 71.06 & 77.00 & 90.21 \\
         Background              & 95.24 & 98.34 & 96.80 \\
         \midrule
         Mean                    & 87.88 & 91.19 & 95.63 \\
         \bottomrule
    \end{tabular}}
    \caption{Results of Mont Blanc Massif Region}
    \label{tab:alps_acc_table}
\end{table}

\begin{figure*}[!t]
    \centering
    
    \begin{subfigure}{0.32\textwidth}
        \includegraphics[width=\columnwidth]{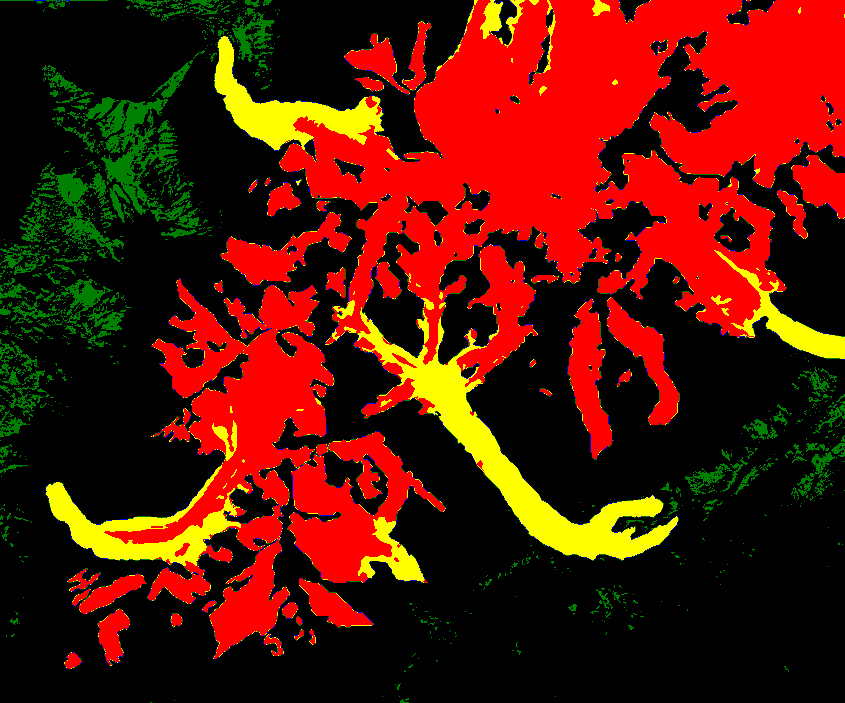}
        \caption{The label mask}
        \label{fig:}
    \end{subfigure}
   \hfill
    \begin{subfigure}{0.32\textwidth}
        \includegraphics[width=\columnwidth]{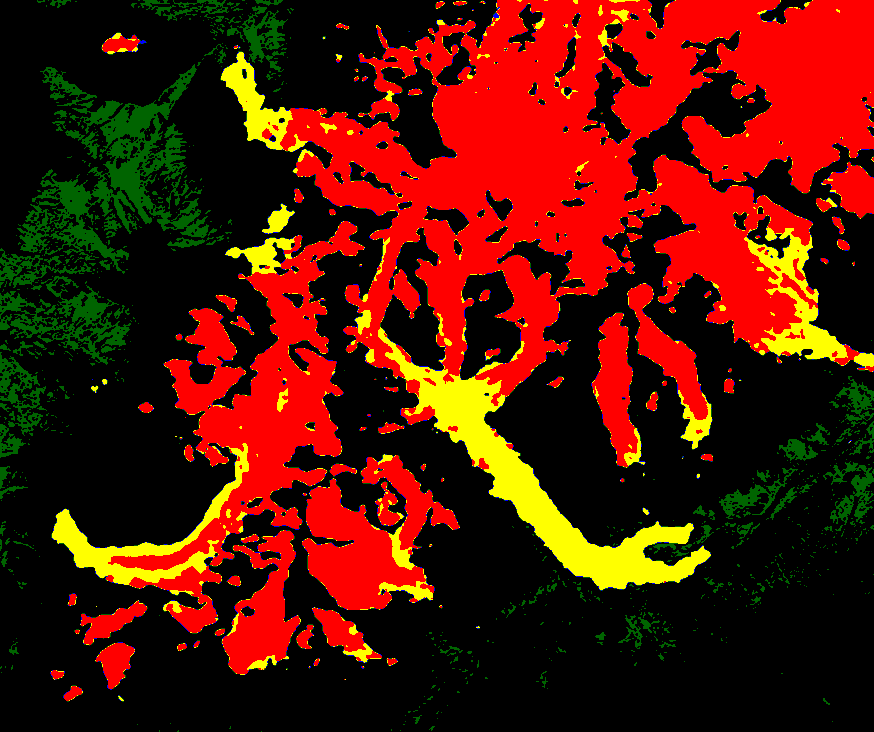}
        \caption{CryoNet}
        \label{fig:}
    \end{subfigure} 
    \hfill
    \begin{subfigure}{0.32\textwidth}
        \includegraphics[width=\columnwidth]{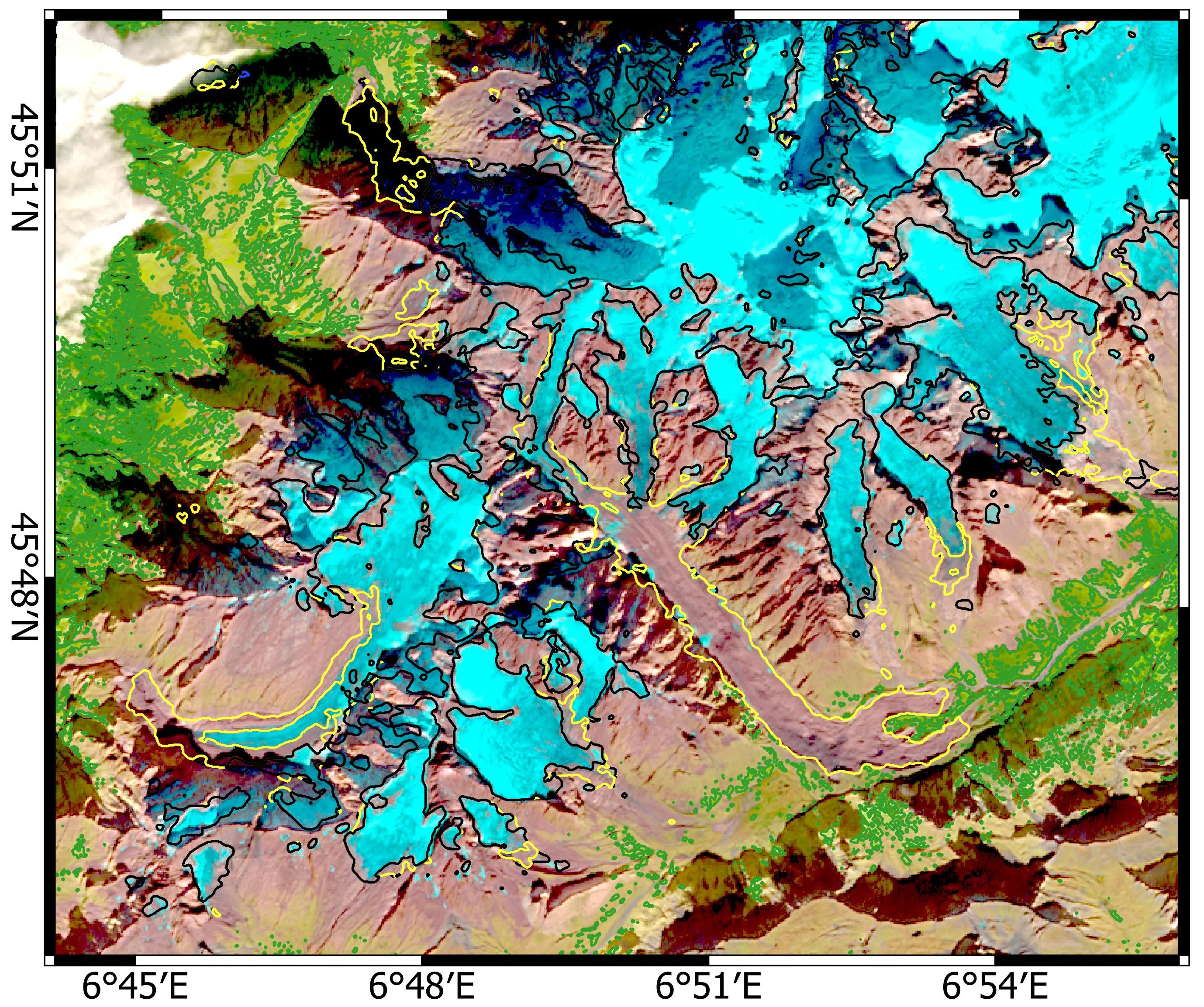}
        \caption{Predictions on Sentinel-2 image}
        \label{fig:}
    \end{subfigure} 
    
    \caption{Prediction results on an unseen subset in Mont Blanc Massif, Alps: (a), (b), (c) }
    \label{fig:Alps_imgs}
\end{figure*}

Fig.\ref{fig:Alps_imgs} illustrates the CryoNet prediction on the Mont Blanc region. The qualitative comparison between the reference labels and the model predictions shows that CryoNet successfully delineates both clean-ice and debris-covered glaciers in this new geographic setting. 

The predicted segmentation maps exhibit a high degree of spatial consistency, preserving the overall glacier extent as well as fine-scale structures such as narrow glacier tongues and boundary transitions. In particular, debris-covered glacier areas are identified with coherent and continuous patterns, despite their spectral similarity to the surrounding terrain. 

Some misclassifications are observed in shadowed regions, including both falsely detected debris-covered areas and missed glacier features, further highlighting the challenges posed by low-illumination conditions.

\begin{figure*}[!t]
\centering

\begin{subfigure}{0.45\textwidth}
    \includegraphics[width=\linewidth]{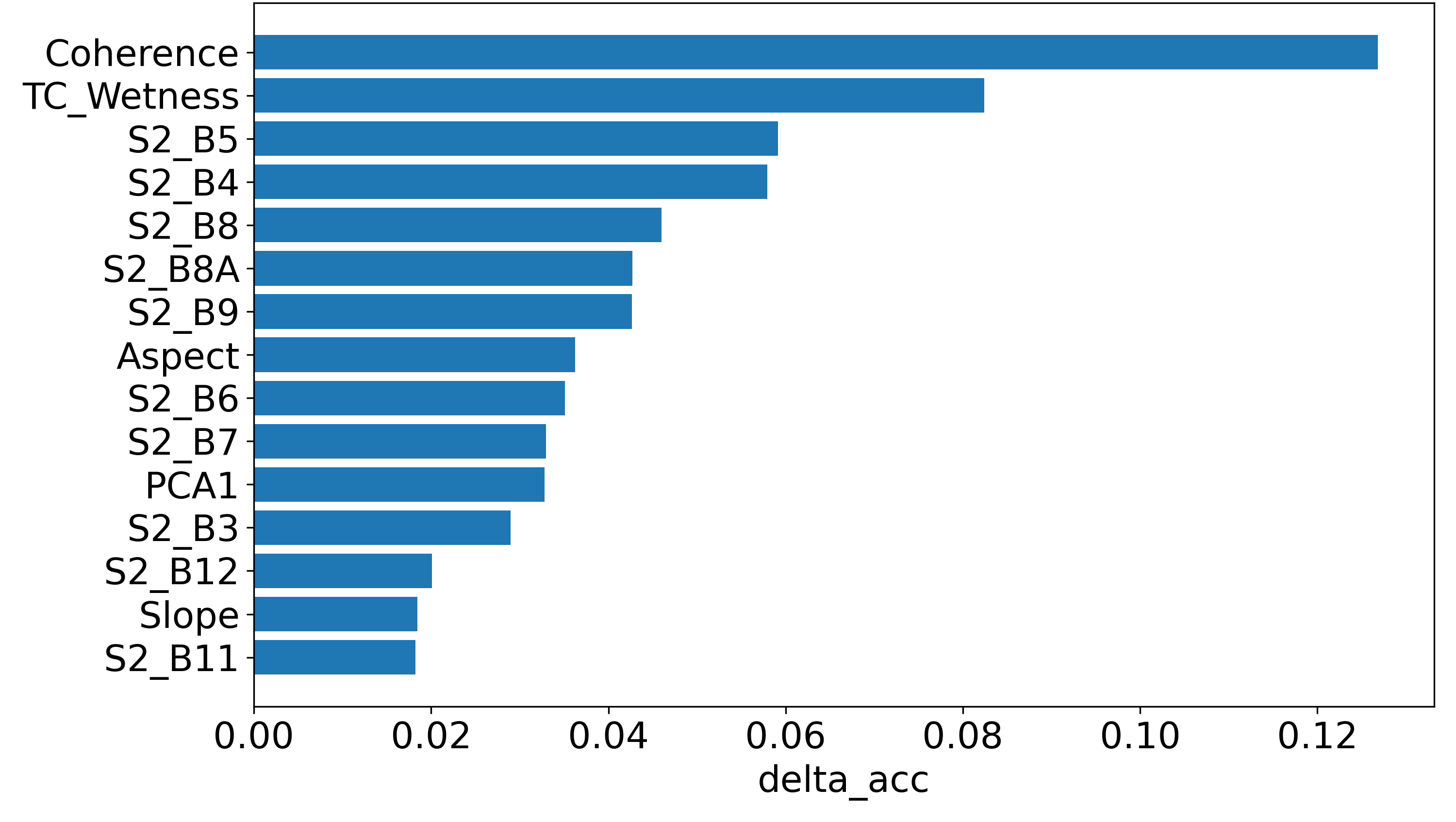}
    \caption{Overall – $\Delta$Accuracy}
    \label{fig:cha_imp_overall_acc}
\end{subfigure}
\hfill
\begin{subfigure}{0.45\textwidth}
    \includegraphics[width=\linewidth]{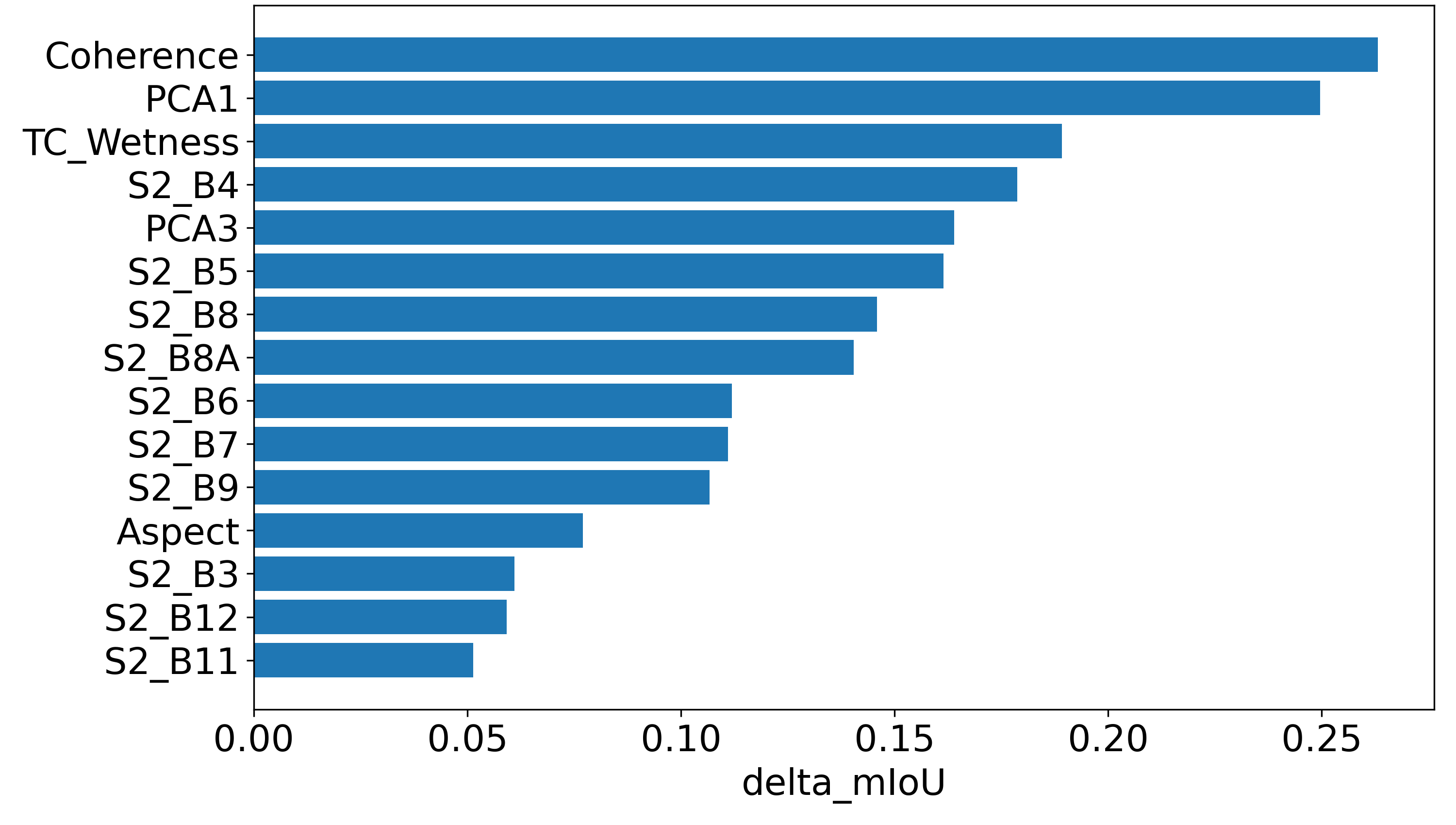}
    \caption{Overall – $\Delta$IoU}
    \label{fig:cha_imp_overall_iou}
\end{subfigure}

\vspace{4mm}

\begin{subfigure}{0.45\textwidth}
    \includegraphics[width=\linewidth]{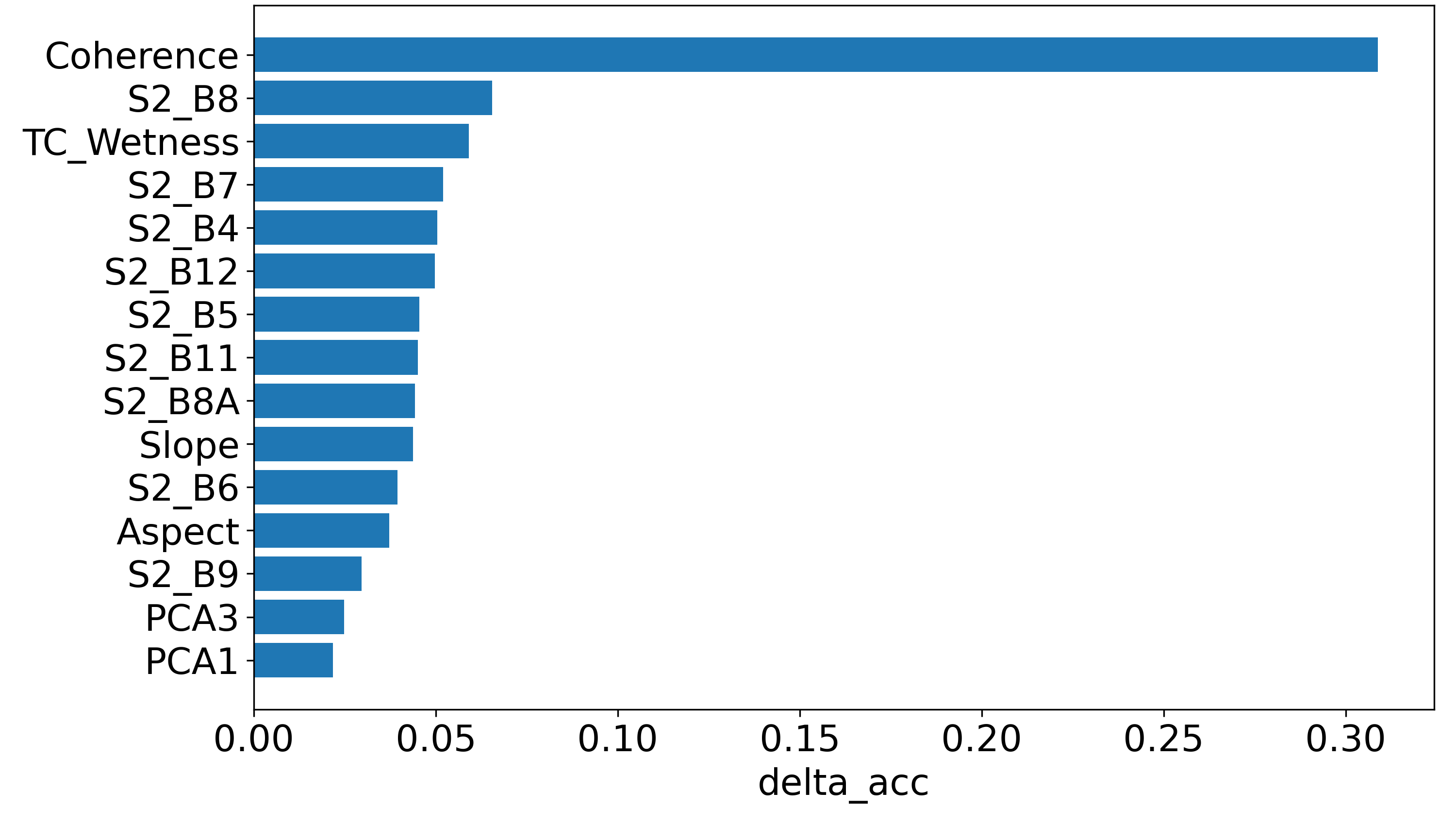}
    \caption{Clean ice – $\Delta$Accuracy}
    \label{fig:cha_imp_clean_acc}
\end{subfigure}
\hfill
\begin{subfigure}{0.45\textwidth}
    \includegraphics[width=\linewidth]{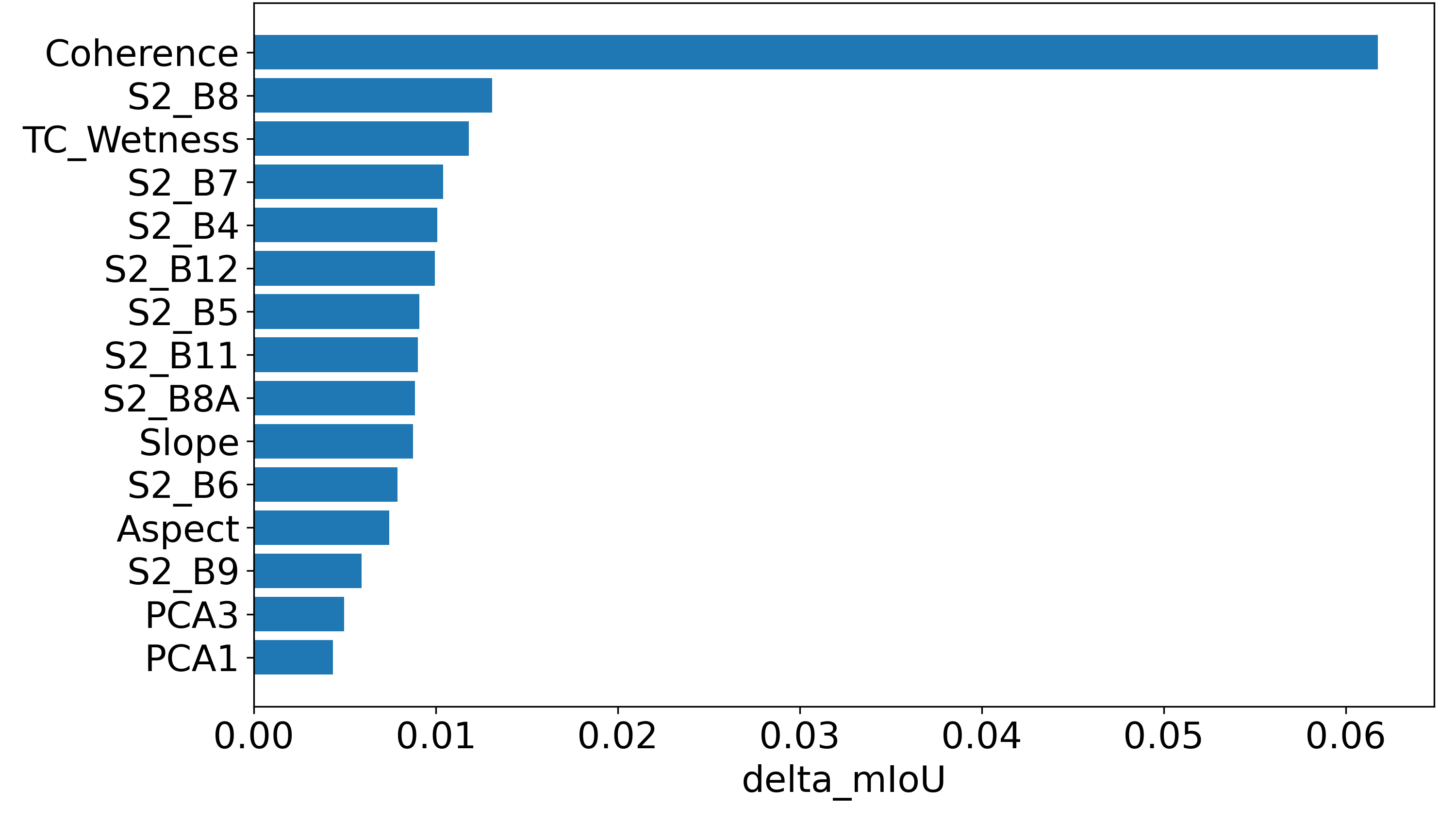}
    \caption{Clean ice – $\Delta$IoU}
    \label{fig:cha_imp_clean_iou}
\end{subfigure}

\vspace{4mm}

\begin{subfigure}{0.45\textwidth}
    \includegraphics[width=\linewidth]{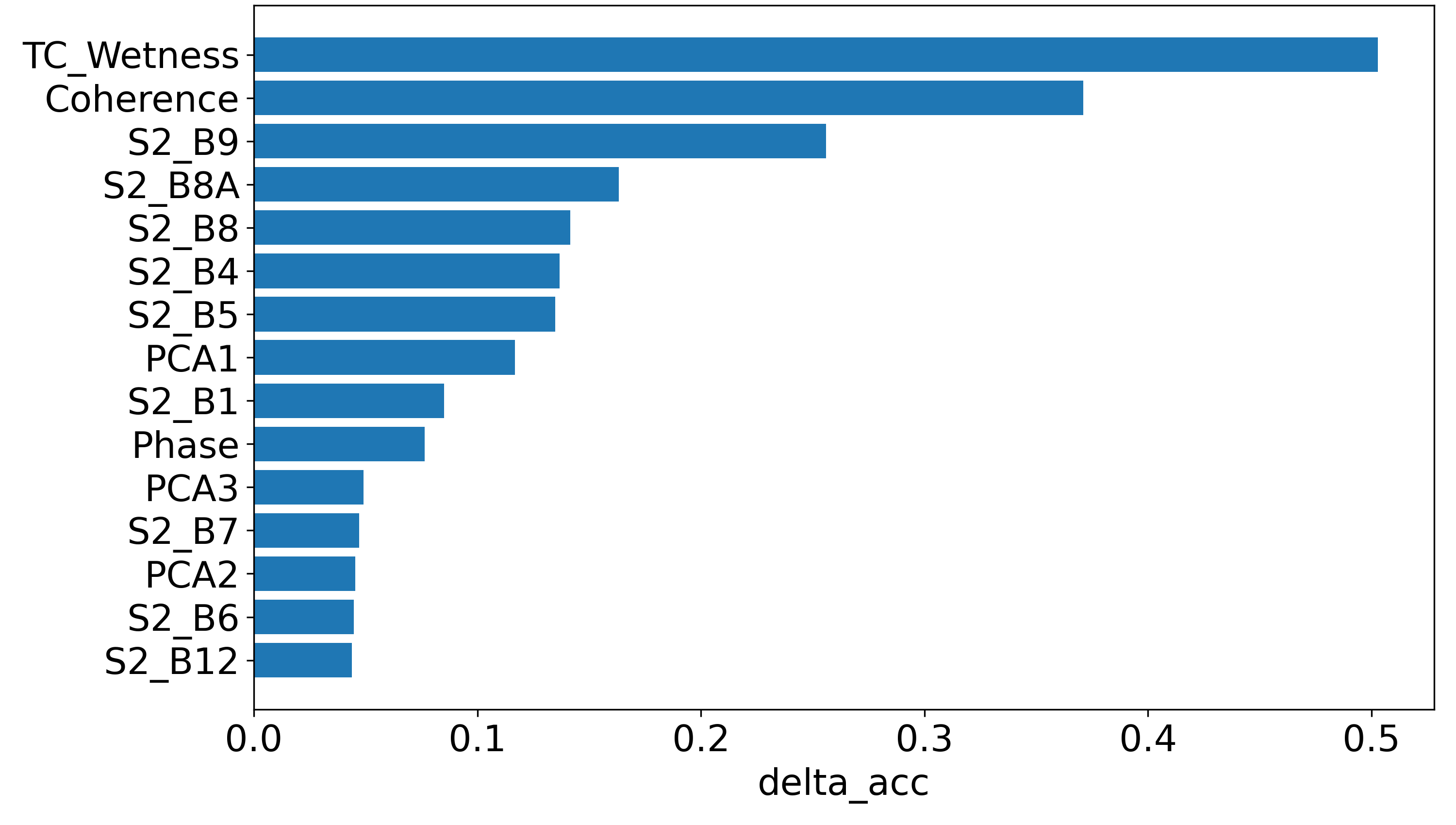}
    \caption{Debris-covered – $\Delta$Accuracy}
    \label{fig:cha_imp_debris_acc}
\end{subfigure}
\hfill
\begin{subfigure}{0.45\textwidth}
    \includegraphics[width=\linewidth]{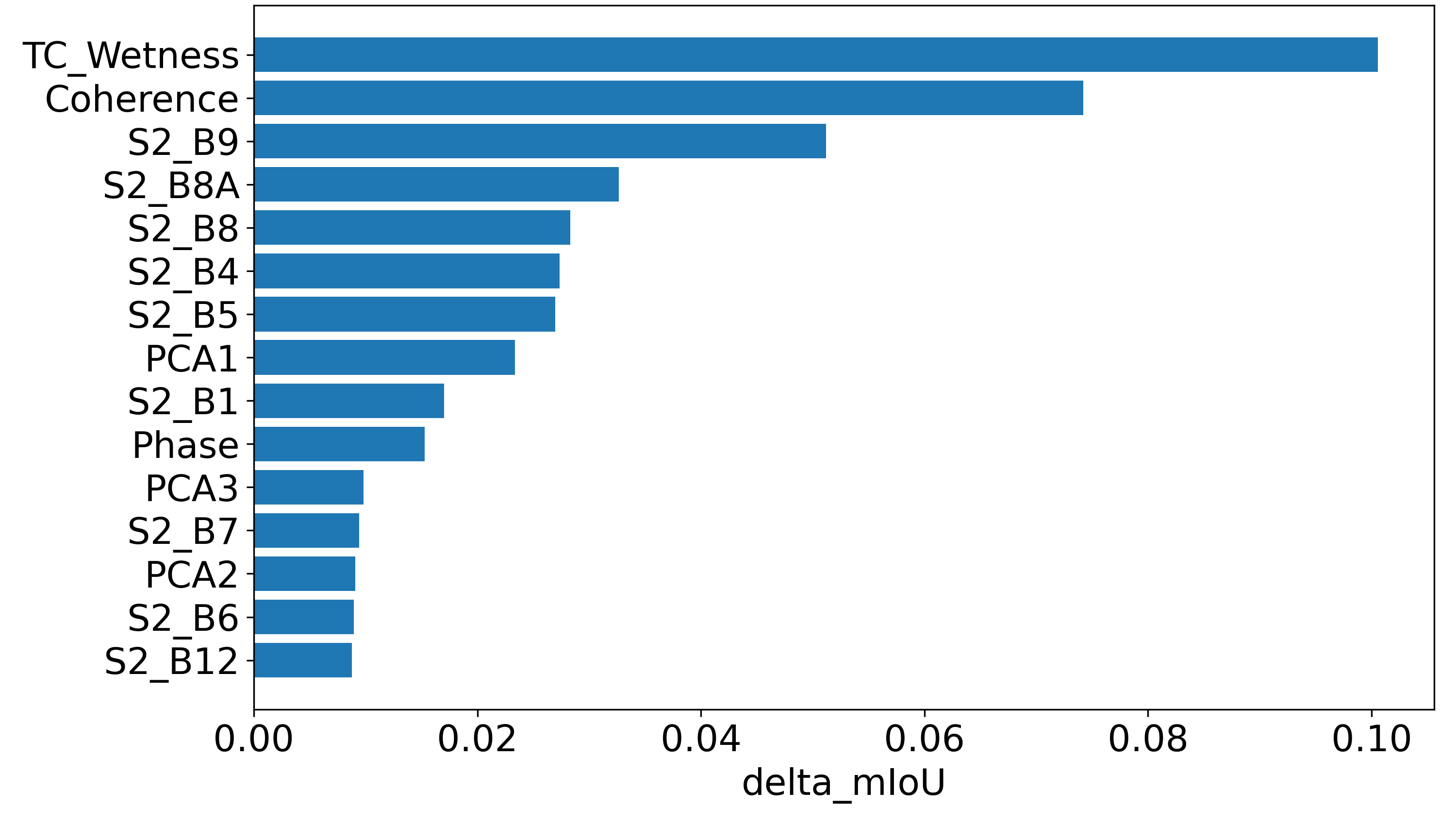}
    \caption{Debris-covered – $\Delta$IoU}
    \label{fig:cha_imp_debris_iou}
\end{subfigure}

\caption{Permutation-based channel importance for overall performance (top row), clean ice mapping (middle row), and debris-covered glacier mapping (bottom row). The left column shows $\Delta$Accuracy, and the right column shows $\Delta$IoU for each band. Larger drops indicate higher band importance.}
\label{fig:cha_imp_all_combined}
\end{figure*}

To further interpret these results and to quantify the role of each data layer in the model's performance, we applied the permutation-based channel importance analysis described in Section~\ref{sec:channel_importance}. Fig.~\ref{fig:cha_imp_all_combined} illustrates the top 15 most influential bands for the overall performance, clean-ice, and debris-covered glacier mapping. For each case, the reduction in accuracy and IoU after permuting individual bands is shown, where larger drops indicate a stronger dependency on that band’s information.

Fig.~\ref{fig:cha_imp_overall_acc} and Fig.~\ref{fig:cha_imp_overall_iou} indicate that the model relies most strongly on SAR coherence, which produces the largest drop in both accuracy and mIoU when perturbed. TC wetness is among the next most influential features, ranking second in accuracy drop and third in mIoU, highlighting the importance of moisture-related information in glacier detection.

PCA components, particularly PCA1, also contribute significantly, especially in terms of mIoU. This is expected, as PCA captures the dominant spectral variability across Sentinel-2 bands, providing compact and informative representations of the scene. In addition, several Sentinel-2 spectral bands, including B4 (red), B5 (red-edge), and B8 (NIR), appear among the most influential features. These bands are well known for their sensitivity to snow, ice, water, and vegetation, which explains their contribution to overall segmentation performance.

As shown in Fig.~\ref{fig:cha_imp_clean_acc} and Fig.~\ref{fig:cha_imp_clean_iou}, SAR coherence remains the most influential feature for clean-ice glacier mapping, confirming its strong capability to distinguish glacier surfaces from stable terrain. Sentinel-2 band B8 (NIR) and TC wetness follow in importance, further emphasizing the role of spectral and moisture-related information. Topographic variables such as slope and aspect also appear among the relevant features, as expected, reflecting the geomorphological characteristics of glacier surfaces.

Figures~\ref{fig:cha_imp_debris_acc} and Fig.~\ref{fig:cha_imp_debris_iou} demonstrate that debris-covered glacier mapping is primarily driven by TC wetness and SAR coherence. This highlights the importance of combining physical and structural information when spectral separability is limited. The results suggest that incorporating TC wetness and coherence features can substantially improve the detection of debris-covered glaciers.

\section{Discussion}
\label{sec:disscussion}
The experimental results demonstrate that integrating multi-source data with the proposed attention-enhanced nested encoder-decoder architecture substantially improves the delineation of debris-covered and clean-ice glaciers. Compared with DeepLabV3+, SegFormer, and the classical U-Net, CryoNet consistently achieved higher accuracy, precision, and IoU across all classes. 

In contrast to existing approaches, CryoNet contributes to addressing several key limitations highlighted in the literature.
First, several studies do not explicitly focus on debris-covered glaciers or do not separate them from clean ice~\cite{Maslov2025, Xiang2025, diaconu2025dl4gam}. 
Second, some methods rely on post-processing pipelines to refine glacier outlines, which increases complexity and reduces automation, as shown in~\cite{Xie2022}.
Third, other studies use limited data modalities; for example,~\cite{Xie2020} relies primarily on Landsat imagery, DEM, and geomorphological indices, or ~\cite{Yang2024MappingAlgorithms}, which incorporates high-resolution optical imagery along with topographic and thermal features. 
Moreover, existing studies generally do not provide a structured evaluation of the relative importance of different data layers, resulting in limited guidance for selecting appropriate data modalities or features in deep learning-based debris-covered glacier mapping.
In this work, we introduce an end-to-end, multi-modal deep learning framework for debris-covered glacier mapping and provide a systematic evaluation of input data contributions.

This CryoNet performance can be attributed to the synergy of several factors: 

\begin{itemize}
    \item The nested skip pathways that promote multiscale feature reuse and preserve spatial consistency.
\end{itemize}

\begin{itemize}
    \item The scSE attention mechanism, which recalibrates features along both spatial and channel dimensions to enhance class separability in heterogeneous terrain.
\end{itemize}

\begin{itemize}
    \item The use of ResNet-101 as a deep encoder that captures rich contextual and topographic information.
\end{itemize}

\begin{itemize}
    \item The inclusion of complementary information to the optical data, such as optical, thermal, SAR coherence, topographic information, texture, and glacier velocity, proved critical for debris-covered glacier mapping using deep learning methods. Combining these diverse inputs enables the network to learn abstract representations that capture not only spectral information but also the physical characteristics of glaciers, including temperature, texture, and slope. 
\end{itemize}

Qualitative analyses of the test data, as shown in Table~\ref{tab:qualitative_test_predict}, further revealed that CryoNet maintains smooth and continuous glacier outlines, particularly in complex areas. 

Some residual confusion is observed between debris-covered and clean-ice glaciers. From a physical perspective, there is usually a gradual transition between these two classes. At the end of the clean ice boundary, there is a thin debris layer, which gradually increases in thickness. Therefore, there is no clear definition of where the debris-covered part of the glacier starts. Considering this fact, these misclassifications likely arise from mixed pixels in the boundary of debris-covered and clean-ice glaciers. Furthermore, inaccuracies in the used reference labels exist that can lead to confusion. For instance, the reference data used were derived from Landsat imagery with a 30 m spatial resolution, whereas this study used Sentinel-2 imagery and resampled all other layers to 10 m. This difference introduces limitations related to spatial resolution and may affect the precise delineation of glacier boundaries. Moreover, the delineation process often involves manual editing or threshold-based methods, both of which may introduce human-induced errors.

Furthermore, the evaluation on an unseen subset of the dataset demonstrates the ability of the network to generalize to data that were not used during training. The model is able to delineate both clean-ice and debris-covered glaciers coherently within the same region.

To further assess transferability, CryoNet was applied to a different region in the Alps (Mont Blanc massif). In this case, the pre-trained model was fine-tuned to maximize the adaptability. The results show that the model can be effectively adapted to new regions.

These findings indicate that while the model generalizes well within the same domain, transfer to regions with different characteristics benefits from domain adaptation. The integration of physical variables, including slope, aspect, coherence, and thermal information, reduces reliance on region-specific spectral signatures and supports the model’s adaptability. Overall, CryoNet shows promising potential for application in diverse mountain regions when combined with lightweight fine-tuning strategies.

The data layer importance analysis provides a comparative assessment of the contribution of the input features. Both debris-covered and clean-ice glacier mapping are strongly influenced by Tasseled Cap (TC) wetness and InSAR coherence. The high importance of SAR coherence is expected, as it is sensitive to surface motion and temporal changes. This result is consistent with previous studies demonstrating the effectiveness of SAR coherence for glacier delineation, regardless of debris coverage, both in traditional approaches~\cite{atwood2010using} and in deep learning-based methods~\cite{Maslov2025, Robson2020, frey2012compilation}.

The contribution of TC wetness reflects the moisture-related characteristics of glacier surfaces and confirms its usefulness for glacier mapping, as also reported in~\cite{Roberts-Pierel2022}. The first PCA component also plays an important role in the overall model performance, highlighting its ability to reduce spectral redundancy and enhance feature representation, consistent with previous studies~\cite{Racoviteanu2012}.

Clean-ice glacier mapping further shows a strong dependence on Sentinel-2 Band 8 (NIR). This is consistent with the sensitivity of the near-infrared region to differences in snow and ice properties, where snow and clean ice generally show relatively high reflectance compared to the surrounding terrain. However, older glacier ice can exhibit lower reflectance than fresh snow. This highlights the usefulness of NIR information for separating clean glacier surfaces from adjacent non-glacier areas. Overall, the layer importance analysis further emphasizes the need to integrate multi-source data for debris-covered glacier mapping using DL methods, as also suggested for future work in~\cite{zhang2026}.

Overall, the data layer importance analysis provides a structured comparison of feature contributions. There are valuable studies such as~\cite{racoviteanu2009challenges} and~\cite{Paul2017}; however, these studies mainly offer general recommendations and do not provide a systematic evaluation of the relative importance of different data sources. As a result, there is still a lack of clear guidance regarding which data layers are most informative for automated glacier mapping, particularly for debris-covered glaciers.

Although CryoNet shows high performance, several limitations remain. A potential source of uncertainty in glacier mapping arises from the presence of seasonal snow that may cover both glacier surfaces and the surrounding terrain at similar elevations, making glacier boundaries difficult to distinguish. To mitigate this effect, we used a scene in which seasonal snow cover is minimal across most of the study area. This selection reduces the likelihood of confusion between transient snow and glacier surfaces, thereby improving the consistency between input data and reference labels. Consequently, the influence of seasonal snow on the model results is expected to be limited. If suitable scenes at the end of the ablation period with minimal snow cover are not available, seasonal snow can obscure glacier boundaries, particularly in the upper parts of glaciers at higher elevations and over small clean-ice glaciers, making them difficult to identify. Future research could deploy multi-temporal inputs to enhance robustness under varying seasonal conditions. Using multiple scenes acquired throughout the year enables the model to learn and discriminate between seasonal snow and glacier ice.

Shadowed regions are another source of uncertainty, as reduced illumination affects spectral information and may lead to misclassification. The issue is further influenced by the acquisition date, as shadows become more pronounced later in the year due to lower solar angles. As observed in the Mont Blanc region, some misclassifications are associated with shadowed areas. In this context, imagery acquired closer to the end of the ablation season (e.g., August or early September) is generally preferable, provided that snow conditions are suitable. Such acquisitions offer a better balance between minimal seasonal snow cover and reduced shadow effects.

Additionally, although multi-modal fusion improved performance, the disparity in spatial resolution among the input datasets introduces potential sources of error. The spatial resolution of the employed data ranges from 10 m for Sentinel-2 to 120 m for the velocity map. The coarse resolution of some data layers may reduce the model’s overall performance.

Finally, an additional source of uncertainty arises from the reference labels. These labels are derived from multiple sources, datasets, and delineation methods, which introduces inconsistencies in both spatial resolution and acquisition time between the labels and the input imagery. As discussed earlier, such discrepancies are a common limitation in glacier mapping, particularly in DL–based approaches when labelled data is required. Despite this, DL models can still achieve robust performance, as they are capable of learning meaningful patterns even in the presence of incomplete or imperfect labels.

Overall, CryoNet demonstrates that this architecture can achieve accuracy comparable to expert delineations in glacier delineation while reducing the need for heuristic post-processing. The framework provides a basis for automated glacier inventory generation that can be scaled for long-term monitoring across diverse mountain regions.

\section{Conclusion}
\label{sec:conclusion}

This study introduced CryoNet, a deep learning framework for automated delineation of debris-covered glaciers using multimodal remote sensing data. By integrating optical, thermal, topographic, SAR-based, and derived features within an attention-enhanced encoder–decoder architecture, the proposed approach effectively addresses the challenge of discriminating debris-covered glaciers from spectrally similar terrain.

The results demonstrate that CryoNet consistently outperforms state-of-the-art architectures, including DeepLabV3+, SegFormer, and U-Net, achieving superior performance across all evaluation metrics. In particular, the model shows significant improvement in mapping debris-covered glaciers, highlighting the effectiveness of combining multi-source data with spatial–channel attention and multi-scale feature fusion. The transferability experiment conducted in the Mont Blanc massif further demonstrates the adaptability of the proposed framework to different geographic settings with limited fine-tuning.

The channel importance analysis further clarifies the role of various features, such as PCA components, InSAR coherence, Tasseled Cap wetness, and topographic features, in accurate glacier delineation using deep learning. By quantitatively assessing the contribution of individual input layers, this study provides practical guidance for selecting informative data sources in automated glacier mapping, particularly for debris-covered glaciers where spectral information alone is insufficient.

Despite the strong performance achieved by CryoNet, several limitations remain. These include sensitivity to seasonal snow and shadow effects, inconsistencies in spatial resolution among input data, and uncertainties associated with multi-source reference labels. Such limitations are typical when using open-access remote sensing datasets, which often differ in spatial resolution, acquisition time, and production methodology. However, one of the main aims of this study was to evaluate the performance of an automated glacier mapping framework using only open-access data. Despite these constraints, the results demonstrate that CryoNet remains robust and provides a reliable framework for automated glacier mapping.

Overall, CryoNet provides a scalable and effective solution for automated glacier mapping, with strong potential to support large-scale monitoring and contribute to improved understanding of glacier dynamics in a changing climate.

\section{Data and Code Availability}
\label{sec:data_availability}

The data and code used in this study will be made publicly available through a GitHub repository upon acceptance of this work.

\balance
\bibliographystyle{IEEEtran.bst}
\bibliography{main}
\end{document}